\documentclass[aps,prc,showpacs,amsmath,showkeys,superscriptaddress]{revtex4}

\usepackage[dvips]{graphicx}

\newcommand{\ba}{\begin{eqnarray}}
\newcommand{\ea}{\end{eqnarray}}
\newcommand{\bmath}{\begin{subequations}}
\newcommand{\emath}{\end{subequations}}
\newcommand{\ban}{\begin{eqnarray*}}
\newcommand{\ean}{\end{eqnarray*}}

\begin{document}

\title{Test of Pseudospin Symmetry in Deformed Nuclei}

\author{J.N. Ginocchio}
 \email{gino@t5.lanl.gov}
 \affiliation{Theoretical Division, Los Alamos National Laboratory,
              Los Alamos, New Mexico 87545, USA}

\author{A. Leviatan}
 \email{ami@vms.huji.ac.il}
 \affiliation{Racah Institute of Physics, The Hebrew University,
              Jerusalem 91904, Israel}

\author{J. Meng}
 \email{mengj@pku.edu.cn}
 \affiliation{School of Physics, Peking University, Beijing 100871, China}

\author{Shan-Gui Zhou}
 \email{sgzhou@mpi-hd.mpg.de}
 \affiliation{Max-Planck Institute for Nuclear Physics, 69029 Heidelberg,
Germany}
 \affiliation{School of Physics, Peking University, Beijing 100871, China}

\date{\today}

\begin{abstract}
Pseudospin symmetry is a relativistic symmetry of the Dirac
Hamiltonian with scalar and vector mean fields equal and opposite
in sign. This symmetry imposes constraints on the Dirac
eigenfunctions. We examine extensively the Dirac eigenfunctions of
realistic relativistic mean field calculations of deformed nuclei
to determine if these eigenfunctions satisfy these pseudospin
symmetry constraints.
\end{abstract}



\maketitle

\section{Introduction}

Pseudospin doublets were introduced more than thirty years ago into
nuclear physics to accommodate an observed near degeneracy of
certain normal parity shell model orbitals with non-relativistic
quantum numbers ($n_r$, $\ell$, $j = \ell + 1/2)$ and ($n_{r}-1,
\ell + 2$, $j = \ell + 3/2$) where $n_r$, $\ell$, and $j$ are the
single-nucleon radial, orbital, and total  angular momentum
quantum numbers, respectively \cite {aa,kth}. The doublet
structure is expressed in terms of a ``pseudo'' orbital angular
momentum, which is an average of the orbital angular momentum of
the two orbits in the doublet, $\tilde{\ell}$ = $\ell$ + 1,
coupled to a ``pseudo'' spin, $\tilde s$ = 1/2 with $j =
\tilde{\ell}\pm \tilde s$. For example, the shell model orbitals
$(n_r s_{1/2},(n_r-1)
d_{3/2})$ will have $\tilde{\ell}= 1$, $(n_r p_{3/2},(n_r-1)
f_{5/2})$ will have $\tilde{\ell}= 2$,
for the two states in the doublet.
Then the single-particle energy is approximately
independent of the orientation of the pseudospin leading to
an approximate pseudospin symmetry. These doublets persist for
deformed nuclei as well \cite{bohr}. The axially-symmetric
deformed single-particle orbits with non-relativistic asymptotic
quantum numbers $[N,n_3,\Lambda]\Omega=\Lambda+1/2$ and
$[N,n_3,\Lambda^{\prime}=\Lambda+2]\Omega^{\prime}=\Lambda+3/2$
are quasi-degenerate.
Here $N$ is the total harmonic oscillator quantum number,
$n_3$ is the number of quanta for oscillations along the symmetry axis,
taken to be in the $z$-direction,
$\Lambda$ and $\Omega$ are respectively the components of the orbital
and total angular momentum projected along the symmetry axis \cite{bm}.
In this case, the doublet structure is expressed in terms of a
``pseudo'' orbital angular momentum projection,
$\tilde{\Lambda} = \Lambda +1$, which is added to a ``pseudo''
spin projection, $\tilde{\mu}=\pm 1/2$ to yield the
above mentioned doublet of states with $\Omega=\tilde{\Lambda} - 1/2$ and
$\Omega^{\prime}=\tilde{\Lambda} + 1/2$.
This approximate pseudospin ``symmetry'' has been used to explain features
of deformed nuclei, including superdeformation \cite{dudek} and
identical bands \cite{twin,stephens,zeng,peter} as well.

Although there have been attempts to understand the origin of this
``symmetry'' \cite{draayer2,draayer}, only recently has it been
shown to arise from  a relativistic symmetry of the Dirac
Hamiltonian \cite {gino,ami} which we review in Section II. This
relativistic symmetry implies conditions on the Dirac
eigenfunctions \cite {gino2} which we discuss in Sections II and
III. These relationships have been studied extensively
\cite{gino2,gino3,ring,meng,ami2,gino4} for spherical nuclei. For
deformed nuclei, the relationships have been studied only in a
limited way and primarily for the lower components of the Dirac
eigenfunctions \cite {tanabe,tanabe1,tanabe2}. In this paper we
shall test thoroughly these relationships between the upper and
lower components of the two states in the doublet for realistic
deformed relativistic eigenfunctions \cite{meng2,meng2a}.

\section{The Dirac Hamiltonian and Pseudospin Symmetry}

The Dirac Hamiltonian, $H$, with an external scalar, $V_S(\vec
r)$, and vector, $V_V(\vec r)$, potentials is given by: \ba H =
\mbox{\boldmath $\hat{\alpha}\cdot p$} + \hat{\beta} \left ( \,M +
V_S(\vec r)\,\right ) + V_V(\vec r)  ~, \label {dirac} \ea where
\mbox{\boldmath $\hat\alpha$}, $\hat\beta $ are the usual Dirac
matrices, $M$ is the nucleon mass, and we have set $\hbar=c=1$.
The Dirac Hamiltonian is invariant under a SU(2) algebra for  two
limits: $V_S(\vec r) = V_V(\vec r)  + C_s$ and  $V_S(\vec r) = -
V_V(\vec r)  + C_{ps}$ where $C_s,C_{ps}$ are constants
\cite{bell}. The former limit has application to the spectrum of
mesons for  which the spin-orbit splitting is small \cite{page}
and for the spectrum of an antinucleon in the mean-field of
nucleons \cite{gino5,meng3}. The latter limit leads to pseudospin
symmetry in nuclei \cite{gino}. This symmetry occurs independent
of the shape of the nucleus: spherical, axial deformed, or
triaxial.

\subsection{Pseudospin Symmetry Generators}

The generators for the pseudospin SU(2) algebra,
${\tilde{S}}_i \;( i = x,y,z)$, which commute with the Dirac Hamiltonian,
$[\,H_{ps}\,,\, {\tilde{S}}_i\,] = 0$, for the pseudospin
symmetry limit $V_S(\vec r)= -V_V(\vec r)+ C_{ps}$, are given by
\cite{ami}
\ba
{{\tilde {S}}}_i =
\left (\begin{array}{cc}
{\tilde s}_i &  0 \\ 0 & { s}_i
\end{array}
\right )
= \left (
\begin{array}{cc}
U_p\, { s}_i \, U_p & 0 \\
0 & { s}_i
\end{array}
\right )
\label{psg}
\ea
where ${ s}_i = \sigma_i/2$ are the usual spin generators,
$\sigma_i$ the Pauli matrices, and $U_p = \, {\mbox{\boldmath
$\sigma\cdot p$} \over p}$ is the momentum-helicity unitary
operator introduced in \cite {draayer}. Thus the operators
${\tilde S}_i$ generate an SU(2) invariant symmetry of $H_{ps}$.
Therefore, each eigenstate of the Dirac Hamiltonian has a  partner
with the same energy,
\begin{equation}
H_{ps}\, \Phi_{{\tilde k},{\tilde \mu}}^{ps}({\vec r}) =
E_{\tilde k}\,\Phi_{{\tilde k},{\tilde
\mu}}^{ps}({\vec r})
\label{eigen}
\end{equation}
where ${\tilde k}$ are the other quantum numbers and ${\tilde \mu} = \pm
{1\over 2}$ is the eigenvalue of
${\tilde S}_z$,
\begin{equation}
{\tilde S}_z\, \Phi_{{\tilde k},{\tilde \mu}}^{ps}({\vec r}) =
{\tilde \mu} \, \Phi_{{\tilde k},{\tilde \mu}}^{ps}({\vec r}) ~.
\label{Sz}
\end{equation}
The eigenstates in the doublet will be connected by the generators
${\tilde S}_{\pm }={\tilde S}_{x}\pm i {\tilde S}_{y}$,
\begin{equation}
{\tilde S}_{\pm }\,  \Phi_{{\tilde k},{\tilde \mu}}^{ps}({\vec r})
= \sqrt{{\left ({1 \over 2}
\mp {\tilde \mu} \right )
\left ({3 \over 2} \pm {\tilde \mu} \right )}}  \,
\Phi_{{\tilde k},{\tilde \mu} \pm 1}^{ps}({\vec r}) ~.
\label{S+}
\end{equation}

The fact that Dirac eigenfunctions belong to the spinor
representation of the pseudospin SU(2), as given in
Eqs.~(\ref{Sz})-(\ref{S+}),
leads to the conditions on the corresponding Dirac amplitudes that
are explored in this paper and developed in the next Subsection.

\subsection{Dirac Eigenfunctions and Pseudospin Symmetry}

An eigenstate $\Phi_{{\tilde k},{\tilde \mu}}^{ps}({\vec r})$ of the Dirac
Hamiltonian $H_{ps}$, Eq.~(\ref{eigen}), is a four-dimensional vector,
\begin{equation}
\Phi_{{\tilde k},{\tilde \mu}}^{ps}({\vec r}) = \left( \begin{array}{c}
g^+_{{\tilde k},{\tilde \mu}}({\vec r}) \\
g^-_{{\tilde k},{\tilde \mu}}({\vec r})\\ if^+_{{\tilde k},{\tilde \mu}}({\vec
r})\\if^-_{{\tilde k},{\tilde \mu}}({\vec r}) \end{array} \right) ~,
\label {four}
\end{equation}
where $g^{\pm}_{{\tilde k},{\tilde \mu}}({\vec r})$ are the ``upper Dirac
components" and
$f^{\pm}_{{\tilde k},{\tilde \mu}}({\vec r})$ are the ``lower Dirac
components". The superscript $+\,(-)$ indicates spin up (spin down).

The connections between the Dirac
eigenstates of the doublet $(\tilde{\mu}=\pm \frac{1}{2})$
resulting from Eqs. (\ref{Sz})-(\ref{S+})
lead to relationships between the Dirac amplitudes in
Eq.~(\ref {four}) \cite{gino2},
\bmath
\ba
f^+_{{\tilde k}, -{1\over2}}({\vec r}) &=&
f^-_{{\tilde k}, {1\over 2}}({\vec r}) = 0 ~,
\label {lowerzero}\\
f^+_{{\tilde k}, {1\over2}}({\vec r}) &=& f^-_{{\tilde k}, -{1\over
2}}({\vec r})
\equiv f_{{\tilde k}}({\vec r}) ~,
\label {lower}\\
g^+_{{\tilde k}, {1\over2}}({\vec r}) &=& -g^-_{{\tilde k}, -{1\over
2}}({\vec r})
\equiv g_{{\tilde k}}({\vec r}) ~,
\label {upper}
\ea
\label {ul}
\emath
and to first order differential equations,
\bmath
\ba
\left (\,
{\partial \over \partial x} - i {\partial \over \partial y}\,
\right )\, g^{-}_{{\tilde k}, { 1\over 2}}({\vec r})
&=&
\left (\,
{\partial \over \partial x} + i {\partial \over \partial y}\, \right)\,
{g}^{+}_{{\tilde k},- { 1\over 2}}({\vec r})  ~,\\
{\partial \over \partial z}\, {g}^{\pm}_{{\tilde k},\mp {1\over 2}}({\vec r})
&=& \pm \left (\, {\partial \over \partial x} \mp
i {\partial \over \partial y}\, \right )
\, g^{\pm}_{{\tilde k},\pm {1\over 2}}({\vec r}) ~.
\label{dif}
\ea
\label{dif2}
\emath
Thus pseudospin symmetry reduces the eight amplitudes for the two
states in the doublet to four amplitudes. In the next Section we
shall discuss these relations in Eqs.~(\ref {ul})-(\ref {dif2}) for
axially deformed nuclei.

\section{Pseudospin symmetry for Axially Deformed nuclei}

If the potentials are axially symmetric, that is, independent of
the azimuthal angle $\phi$, $V_{S,V}(\vec r) = V_{S,V}(\rho,z)$, $\rho =
\sqrt{x^2 + y^2} $, then the Dirac Hamiltonian has an additional U(1)
symmetry in the pseudospin limit. The conserved $U(1)$ generator
is given by \cite{ami}
\begin{equation}
{ {\tilde L}_z} = \left( \begin{array}{cc}
{ {\tilde \ell}_z} & 0 \\
0 & {\ell_z}  \end{array}
           \right) ~.
\label{Lps}
\end{equation}
where ${\tilde \ell}_z = U_p {\ell}_zU_p$ and
${\ell_z} = ({\bf{r\times p}})_{z}$.
In this case, the Dirac eigenstates of $H_{ps}$ are simultaneous
eigenstates of ${{\tilde L}}_z$,
with eigenvalue ${\tilde\Lambda}$, and of the total angular momentum
generator $J_z = {\tilde S}_z + {\tilde L}_z$,
with eigenvalues $\Omega = {\tilde\Lambda}+\tilde{\mu}
= {\tilde\Lambda}\pm \frac{1}{2}$
\ba
{ {\tilde L}}_z\, \Phi_{{\tilde \eta},{\tilde \Lambda},
{\tilde\mu},\Omega}^{ps}({\vec r})
&=& {\tilde \Lambda}\,
\Phi_{{\tilde \eta},{\tilde \Lambda},{\tilde\mu},\Omega}^{ps}({\vec r}) ~,
\nonumber\\
{J}_z\,\Phi_{{\tilde \eta},{\tilde \Lambda},
{\tilde\mu},\Omega }^{ps}({\vec r})
&=&\Omega\,
\Phi_{{\tilde \eta},{\tilde \Lambda},{\tilde\mu},\Omega}^{ps}({\vec r}) ~.
\label{pLz}
\ea
Here ${\tilde \eta}$ denotes additional quantum numbers that may be needed
to specify the states uniquely.

The conventional method of labeling the eigenstates of axially
deformed single-particle states in nuclei is to use the
asymptotic quantum numbers $(N,n_3,\Lambda,\Omega)$, mentioned
in the Introduction, that emerge in the limit of a non-relativistic
axially-symmetric deformed harmonic oscillator with spin symmetry.
For the relativistic axially-deformed harmonic oscillator with spin
symmetry \cite{gino10} the eigenfunctions
can also be labeled by these quantum numbers. However, only the spatial
amplitudes of the upper components
of the doublet will necessarily have the nodes suggested by these
quantum numbers, whereas the spatial amplitudes of the lower components
may have different nodal structure.
For spherically symmetric potentials a general
theorem relates the nodal structure of the upper and lower
Dirac amplitudes, and has been used to explain
the non-relativistic radial quantum numbers characterizing
pseudospin doublets in spherical nuclei \cite{ami3}.
A corresponding theorem
for axially-deformed potentials in the pseudospin and spin limits
of
the Dirac Hamiltonian appears to hold under certain conditions
which
the relativistic harmonic oscillator satisfies, but which
do
not generally apply for realistic axially-symmetric
potentials \cite{ami4}. For the latter, only the quantum
numbers ${\tilde {\Lambda}}$
and $\Omega$ in Eq.~(\ref{pLz})
are conserved in the pseudospin
limit.
The fact that the axial-symmetry of
the potentials determines
the $\phi$-dependence of the Dirac wave functions,
leads to the
following form for the relativistic pseudospin
doublet eigenstates
\cite{gino2}
\bmath
\begin{equation}
\Phi^{ps}_{{\tilde \eta},{\tilde
\Lambda}, -{1\over 2},
\Omega = \tilde \Lambda -{1\over2}}({\vec
r})
= \left(
\begin{array}{c} g^{+}_{{\tilde \eta},{\tilde \Lambda},
-\frac{1}{2}}(\rho, z)\, e^{i  ({\tilde \Lambda} - 1)  \phi} \\
-g_{{\tilde \eta},{\tilde \Lambda}}(\rho, z)\,
e^{i{\tilde \Lambda}\phi}\\
0\\
if_{{\tilde \eta},{\tilde \Lambda}}(\rho, z)\,
e^{i{\tilde \Lambda} \phi}
\end{array}     \right), \quad
\Omega = \tilde \Lambda -{1\over2} ~,
\label{psdd}\\
\end{equation}
\begin{equation}
\Phi^{ps}_{{\tilde \eta},{\tilde \Lambda},{1\over2},
\Omega^{\prime} = \tilde \Lambda +{1\over2}}({\vec r}) =
\left(
\begin{array}{c}
g_{{\tilde \eta},{\tilde \Lambda}}(\rho,z)\,
e^{i  {\tilde \Lambda} \phi} \\
g^{-}_{{\tilde \eta},{\tilde \Lambda},\frac{1}{2}}(\rho, z)\
e^{i ( {\tilde \Lambda} + 1) \phi}\\
if_{{\tilde \eta},{\tilde \Lambda}}(\rho, z)\,
e^{i  {\tilde \Lambda} \phi}\\0
\end{array} \right), \quad
\Omega^{\prime} = \tilde \Lambda +{1\over2} ~.
\label{psdu}\\
\end{equation}
\label{psd}
\emath

The two states in the doublet have the same pseudo-orbital
angular momentum projection along the symmetry axis, ${\tilde \Lambda}$,
but different total angular momentum projections
$\Omega = {\tilde \Lambda} - {1 \over 2}$ and
$\Omega^{\prime} = {\tilde \Lambda} + {1 \over 2}$.
As seen from Eqs.~(\ref{psdd}) and (\ref{psdu}),
the pseudospin projection, $\tilde{\mu}=\pm\frac{1}{2}$, and
$\tilde{\Lambda}$ are respectively the ordinary spin projection
and ordinary orbital angular momentum projection of the non-vanishing
lower component $f_{{\tilde \eta},{\tilde \Lambda}}(\rho, z)$.
The corresponding dominant upper components
$g^{+}_{{\tilde \eta},\Lambda,-\frac{1}{2}}(\rho, z)$
and $g^{-}_{{\tilde \eta},\Lambda,\frac{1}{2}}(\rho, z)$
have orbital angular momentum projections
$\Lambda = \tilde{\Lambda}-1$ and
$\Lambda^{\prime} = \tilde{\Lambda}+1$ respectively,
hence $\Lambda^{\prime}=\Lambda+2$. Accordingly,
$\Omega=\Lambda+1/2$ and
$\Omega^{\prime}={\Lambda}^{\prime}-1/2=\Lambda+3/2$.
These assignments agree
with the non-relativistic pseudospin quantum numbers
discussed in the Introduction.
The generic label ${\tilde \eta}$ in
$\Phi^{ps}_{{\tilde
\eta},\tilde{\Lambda},\tilde{\mu},\Omega}(\vec{r})$ replaces
the harmonic oscillator labels $N$ and $n_3$, which are not conserved
for realistic axially-deformed potentials in nuclei.

In obtaining the expressions in Eq.~(\ref{psd}) we  have used the
relations in Eq.~(\ref{ul}), which for axially-deformed
potentials
read \cite{gino2}
\bmath
\ba
f^+_{{\tilde \eta},
{\tilde\Lambda},-{1\over2}}(\rho,z) &=&
f^-_{{\tilde \eta},
{\tilde\Lambda},{1\over2}}(\rho,z) = 0 ~,
\label
{psc0}\\
f^+_{{\tilde \eta}, {\tilde
\Lambda},{1\over2}}(\rho,z)
&=&
f^-_{{\tilde \eta}, {\tilde\Lambda},-{1\over2}}(\rho,z)
\equiv
f_{{\tilde \eta}, {\tilde\Lambda}}(\rho,z) ~,
\label
{pscf}\\
g^+_{{\tilde \eta}, {\tilde\Lambda},{1\over2}}(\rho,z)
&=&
-  g^-_{{\tilde \eta}, {\tilde\Lambda},-{1\over2}}(\rho,z)
\equiv
g_{{\tilde \eta},{\tilde\Lambda}}(\rho,z) ~,
\label {pscg}
\ea
\label{psc}
\emath
and the differential relations (\ref {dif2})
become
\bmath
\ba
\left (\,
{\partial \over \partial {\rho}} +
{{\tilde \Lambda} + 1\over \rho}
\,\right )
\, g^{-}_{{{\tilde \eta},
{\tilde \Lambda}, {1\over2}}}(\rho, z)
&=&
\left (\,{\partial
\over
\partial {\rho}} - {{\tilde \Lambda} - 1\over \rho}\,
\right
)\, g^{+}_{{\tilde \eta},{\tilde \Lambda},- {1\over2}}(\rho, z)
~,
\label{eq:diff1}\\
{\partial \over \partial {z}}\,
g^{\pm}_{{\tilde \eta},
{\tilde \Lambda},\mp
{1\over2}}(\rho,
z)
&=&
\pm \left (\,{\partial \over \partial {\rho}} \pm
{{\tilde
\Lambda}\over \rho}\, \right )
\, g^{\pm}_{{\tilde \eta},
{\tilde \Lambda},\pm {1\over2}}(\rho, z)
~.
\label{eq:diff2}
\ea
\label{psdiff}
\emath
We shall now test to
see if the pseudospin symmetry conditions
in Eqs.
(\ref{psc})-(\ref{psdiff}) are valid for realistic relativistic
mean
field eigenfunctions in deformed nuclei.

\section{Comparison with Realistic Relativistic Eigenfunctions}

The single-particle energies and wave functions for $^{168}$Er are
calculated by the relativistic Hartree theory with the parameter
set NL3 in a Woods-Saxon basis \cite{meng2,meng2a}. This method
has been developed from relativistic theory in coordinate space
\cite{meng96prl,meng98npa,meng98prl} and has the advantage that it
easily generalizes to include both deformation and pairing
correction self-consistently. The pairing correlation is treated
with the BCS approximation. These calculations lead to a
theoretical average binding energy $B/A = 8.107$ MeV, a quadrupole
deformation $\beta=0.3497$ and a root mean square radius $R=5.376$
fm, which reproduce the data well. For these realistic
eigenfunctions the harmonic oscillator quantum numbers are not
conserved, but the orbitals are labeled by the quantum numbers of
the main spherical basis state in the expansion of the dominant
upper component in the Dirac eigenfunctions.

In Figure 1, the calculated single-neutron energies, $\varepsilon
= E - M$, for the pseudospin doublets in $^{168}$Er are presented.
From left to right, the panels correspond to the pseudo orbital
angular momentum projection $\tilde{\Lambda} =$ 1, 2, 3 and 4,
respectively. The energy splitting between members of pseudospin
doublets decreases as the single-particle binding energy,
-$\varepsilon$, decreases. For pseudospin doublets with binding
energy larger than 5 MeV, the spin-up (pseudospin down) state is
higher than the spin-down (pseudospin up) one. On the other hand,
for the bound doublets with binding energy less than 5 MeV, the
opposite is observed.

Four pairs of neutron pseudospin partners are chosen to illustrate
the relations given above. (i)~The states $[402]{5\over2}$ and
$[404]{7\over2}$ ($\tilde{\Lambda}=3$), which have a large energy
splitting (about 2 MeV). The single-particle energies are
respectively $\varepsilon_{[402]{5\over2}}=-12.083$ and
$\varepsilon_{[404]{7\over2}}=-14.160$ MeV. (ii)~The states
$[400]{1\over2}$ and $[402]{3\over2}$ ($\tilde{\Lambda}=1$), which
have a small energy splitting (about 0.4 MeV). The single-particle
energies are respectively $\varepsilon_{[400]{1\over2}}=-10.2073$
and $\varepsilon_{[404]{7\over2}}=-10.603$ MeV. (iii)~The states
$[501]{3\over 2}$ and $[503]{5 \over 2 }$ ($\tilde{\Lambda}=2$),
which have a small energy splitting (less than 0.4 MeV). The
single-particle energies are respectively
$\varepsilon_{[501]{3\over 2 }}=-1.349$ and $\varepsilon_{[503]{5
\over 2 }}=-0.9603$ MeV. (iv)~The states $[510]{1\over 2}$ and
$[512]{3\over 2}$ ($\tilde{\Lambda}=1$), which have a tiny energy
splitting (less than 0.01 MeV). The single-particle energies are
respectively $\varepsilon_{[510]{1\over 2}}=-3.8436$ and
$\varepsilon_{[512]{3\over 2}}=-3.8378$ MeV.

Plots for the above four pairs of neutron pseudospin partners are
shown in Figures 2, 4, 6, 8, as a function of $z$ for three
segments: $\rho=1,\, 3,\, 5$ fm, and in Figures 3, 5, 7, 9, as a
function of $\rho$ for three segments: $z=1,\, 3,\, 5 $ fm. In
each segment the top row displays the relationships between lower
component amplitudes given in Eqs.~(\ref{psc0})-(\ref{pscf}), and
the relationship between upper component amplitudes given in
Eq.~(\ref{pscg}). The bottom row displays the differential
relationships between upper component amplitudes given in
Eq.~(\ref{psdiff}).

From these figures, we can draw a number of conclusions. First,
while the amplitudes $f^{+}_ {{\tilde \eta},{\tilde\Lambda},-{1
\over 2}}(\rho, z)$, $f^{-}_{{\tilde \eta},{\tilde \Lambda},{1
\over 2}}(\rho, z)$ are not zero as predicted by Eq.~(\ref{psc0}),
they are much smaller than $f^{-}_{{\tilde \eta}, {\tilde
\Lambda},-{1 \over 2}}(\rho, z)$, $f^{+}_{{\tilde \eta},{\tilde
\Lambda},{1 \over 2}}(\rho, z)$. Furthermore, $f^+_{{\tilde \eta},
{\tilde \Lambda},{1\over 2}}(\rho,z)$ and $f^-_{{\tilde
\eta},{\tilde\Lambda},-{1 \over 2}}(\rho, z)$ have similar shapes
as predicted by Eq.~(\ref{pscf}). In Figures 2-3 for the
$[402]5/2,\,[404]7/2,\,{{\tilde \Lambda} = 3}$ doublet there is
some discrepancy in the shapes but the shapes become more equal as
both the pseudo-orbital angular momentum projection ${\tilde
\Lambda}$ decreases (see Figures 4-5 for the
$[400]1/2,\,[402]3/2,\,{{\tilde \Lambda} = 1}$ doublet) and the
binding energy decreases (see Figures 6-7 for the $[501]3/2,\,
[503]\,5/2,\, {\tilde \Lambda}= 2$ doublet).

The amplitude $-g^-_{{\tilde \eta}, {\tilde\Lambda},-{1 \over
2}}(\rho, z)$ has the same shape as the amplitude $g^+_{{\tilde
\eta},{\tilde \Lambda},{1 \over 2}}(\rho, z)$, in line with the
prediction of Eq.~(\ref{pscg}), but they differ in magnitude.
Again the discrepancy decreases as the pseudo-orbital angular
momentum projection ${\tilde \Lambda}$ decreases (compare Figures
2-3 for the $[402]5/2,\,[404]7/2,\,{{\tilde \Lambda} = 3}$ doublet
with Figures 8-9 for the $[510]1/2,\,[512]3/2,\,{{\tilde \Lambda}
= 1}$ doublet) and the binding energy decreases (compare Figures
2-3 for the $[402]5/2,\,[404]7/2,\,{{\tilde \Lambda} = 3}$ doublet
with Figure 6-7 for the $[501]3/2,\,[503]5/2,\,{{\tilde \Lambda} =
2}$ doublet or Figures 8-9 for the $[510]1/2,\,[512]3/2,\,{{\tilde
\Lambda} = 1}$ doublet). These amplitudes are much smaller than
the other upper amplitudes, $g^\pm_{{\tilde \eta},{\tilde
\Lambda},\mp {1\over 2}}(\rho, z)$.

The differential relation in Eq.~(\ref {eq:diff1}) between the
dominant upper components, $g^{-}_{{\tilde
\eta},\tilde{\Lambda},\frac{1}{2}}(\rho,z)$ and $g^{+}_{{\tilde
\eta},\tilde{\Lambda},-\frac{1}{2}}(\rho,z)$, is well obeyed in
all cases. The differential relations in Eq.~(\ref {eq:diff2})
relate the dominant upper components, $g^{\pm}_{{\tilde
\eta},\tilde{\Lambda},\mp\frac{1}{2}}(\rho,z)$ to the small upper
components $g^{\pm}_{{\tilde
\eta},\tilde{\Lambda},\pm\frac{1}{2}}(\rho,z)$. The shapes of the
left-hand-side and of the right-hand-side of Eq.~(\ref {eq:diff2})
are the same, but the corresponding amplitudes are quite
different. Therefore, the differential relations in Eq.~(\ref
{eq:diff2}) are less satisfied. These differences might partly
originate from the differences in the magnitudes of the small
upper components in Eq.~(\ref{pscg}).

\section{Summary }

We have reviewed the conditions that pseudospin symmetry places on
the Dirac eigenfunctions. We have shown that the conditions on the
lower amplitudes, Eqs.~(\ref{psc0})-(\ref{pscf}), are
approximately satisfied for axially deformed nuclei. The
differential relation between the dominant upper component
amplitudes, Eq.~(\ref{eq:diff1}), is also approximately satisfied.
However, both the relation between the amplitudes of the small
upper components, Eq.~(\ref{pscg}), and the differential
equations, Eq.~(\ref{eq:diff2}), that relate the dominant upper
components with the small upper components are not well satisfied.
The pseudospin symmetry improves as the binding energy and
pseudo-orbital angular momentum projection decrease, which is
consistent with previous tests of pseudospin symmetry in spherical
nuclei.

This research was supported in part by the United States
Department of Energy under contract W-7405-ENG-36, in part by a
grant from the US-Israel Binational Science Foundation and in part
by the Major State Basic Research Development Program Under
Contract Number G2000077407 and the National Natural Science
Foundation of China under Grant No. 10025522, 10221003, 10047001
and 19935030.

\renewcommand{\baselinestretch}{1}

\begin{figure}
\includegraphics[width=15cm]{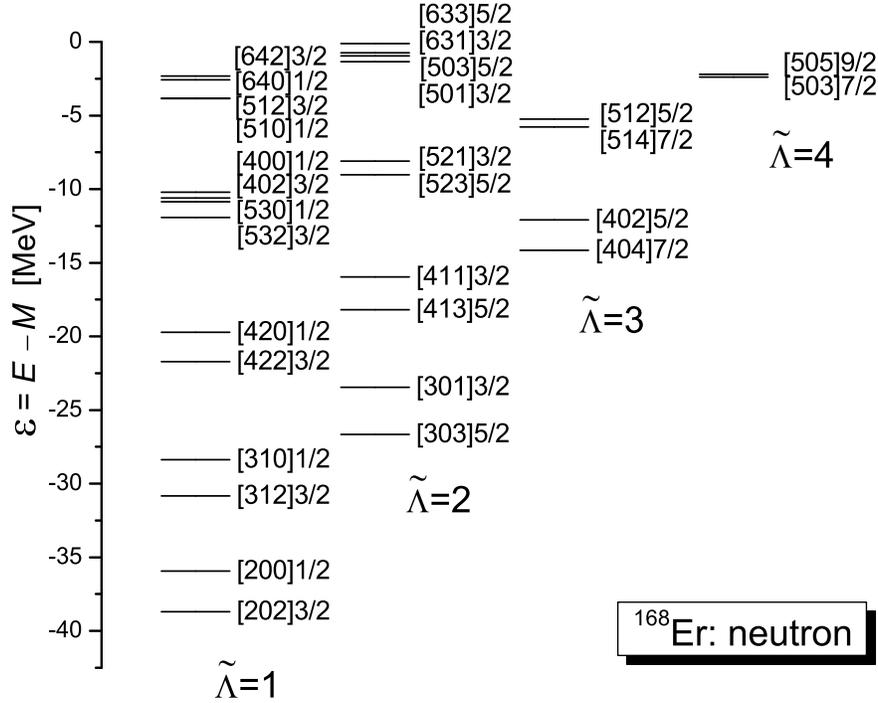}
\caption{The neutron spectra in MeV for the pseudospin doublets in
$^{168}$Er.}\label{fig1}
\end{figure}
\newpage

\begin{figure}
\begin{center}
\includegraphics[width=14.5cm]{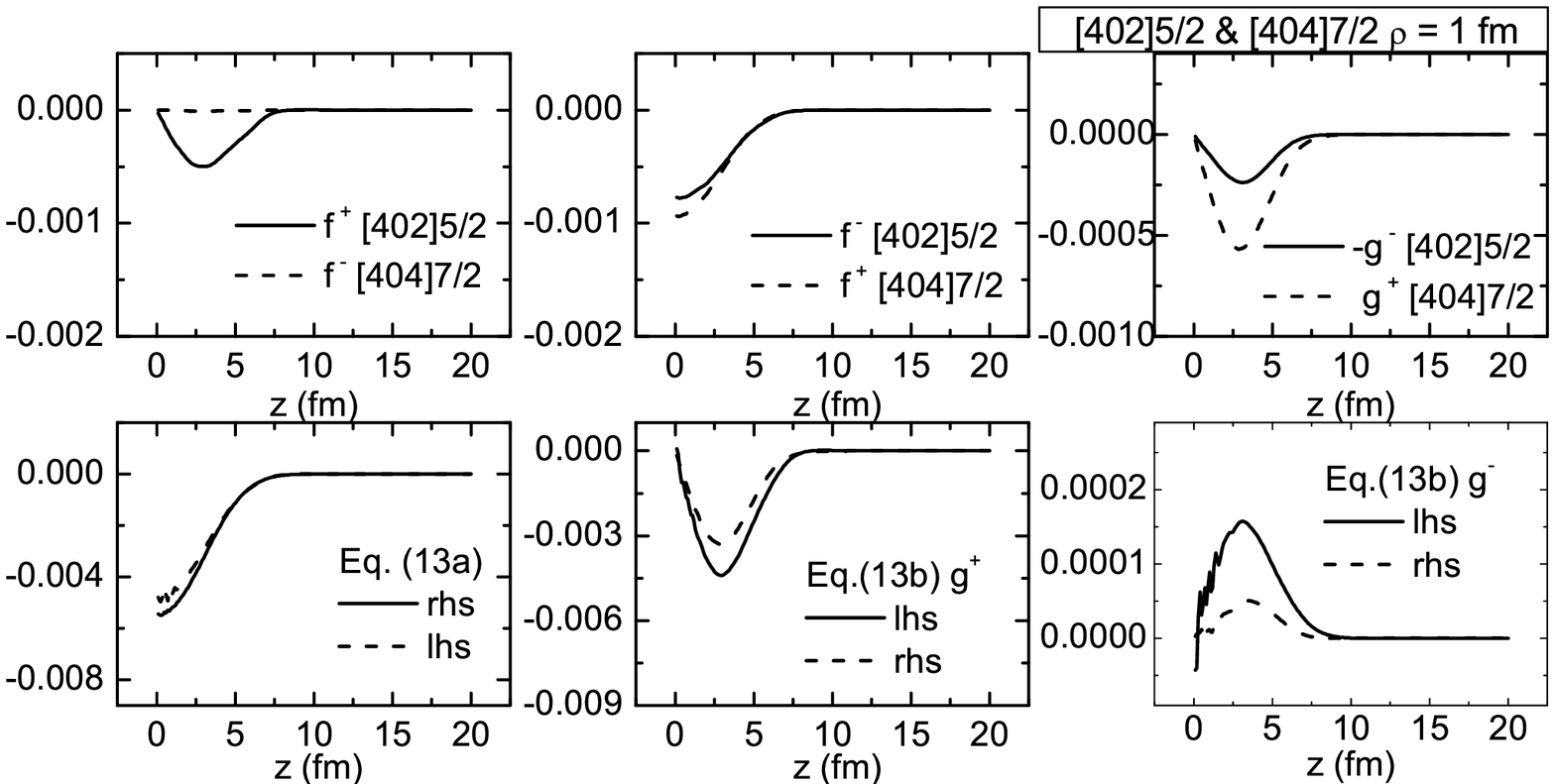}
\includegraphics[width=14.5cm]{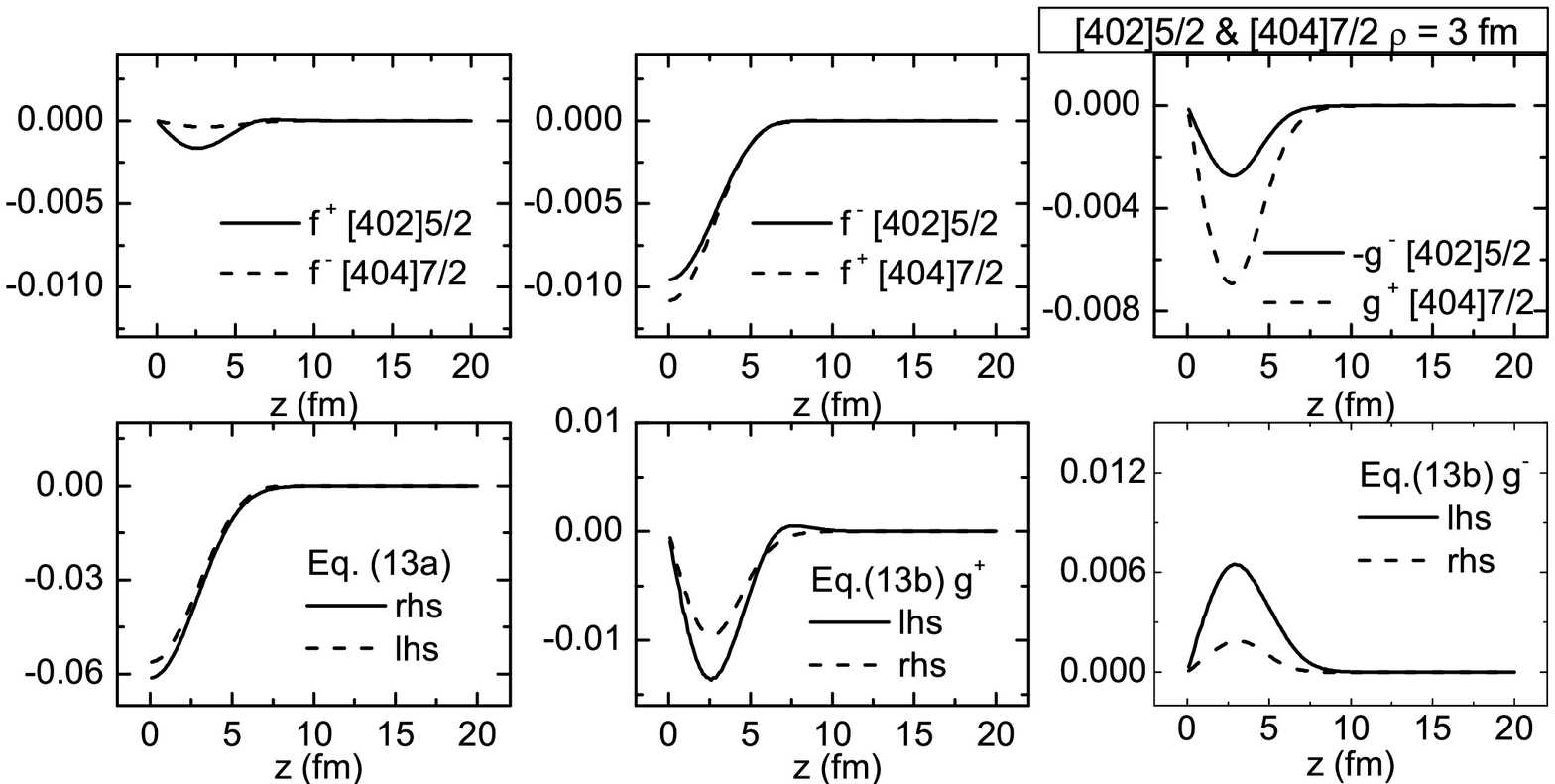}
\includegraphics[width=14.5cm]{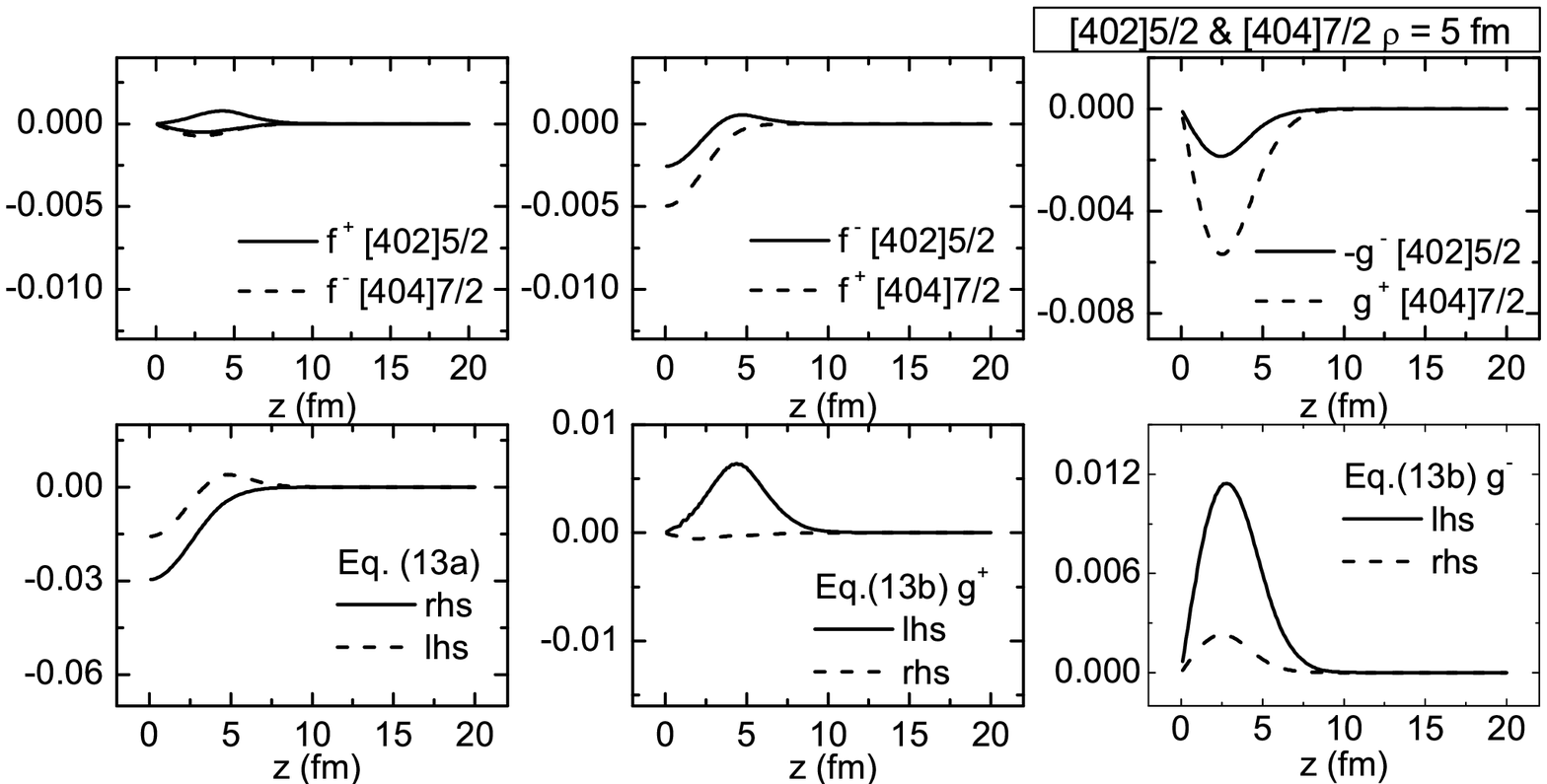}
\end{center}
\vspace{-0.5cm}
\caption{\small
Wave functions in (Fermi)$^{-3/2}$ as a function of $z$ and
$\rho = 1,\, 3,\, 5$ fm for the neutron
pseudospin doublet [402]5/2 and [404]7/2 $({\tilde \Lambda} =3)$
in $^{168}$Er. In each segment, the top row shows (from left to right)
the relations in
(i) Eq.~(\ref{psc0}), involving
$f^{+}_{{\tilde \eta}, {\tilde \Lambda},-1/2}$ and
$f^{-}_{{\tilde \eta}, {\tilde\Lambda},1/2}$,
(ii) Eq.~(\ref{pscf}), involving
$f^{-}_{{\tilde \eta},{\tilde \Lambda},-1/2}$ and
$f^{+}_{{\tilde \eta},{\tilde \Lambda},1/2}$,
(iii) Eq.~(\ref{pscg}), involving
$g^{+}_{{\tilde \eta}, {\tilde \Lambda},1/2}$ and
$-g^{-}_{{\tilde \eta},{\tilde\Lambda},-1/2}$.
The bottom row shows (from left to right) the lhs and rhs of
(i)~Eq.~(\ref{eq:diff1}), involving
$g^{-}_{{\tilde \eta},{\tilde \Lambda},{1\over2}}$
and
$g^{+}_{{\tilde \eta},{\tilde \Lambda},-{1\over2}}$,
(ii)~Eq.~(\ref{eq:diff2}), involving
$g^{+}_{{\tilde \eta},{\tilde \Lambda},-{1\over2}}$
and
$g^{+}_{{\tilde \eta},{\tilde \Lambda},+{1\over2}}$,
(iii)~Eq.~(\ref{eq:diff2}), involving
$g^{-}_{{\tilde \eta},{\tilde \Lambda},+{1\over2}}$
and
$g^{-}_{{\tilde \eta},{\tilde \Lambda},-{1\over2}}$.}
\label{fig2}
\end{figure}

\begin{figure}
\begin{center}
\includegraphics[width=14.5cm]{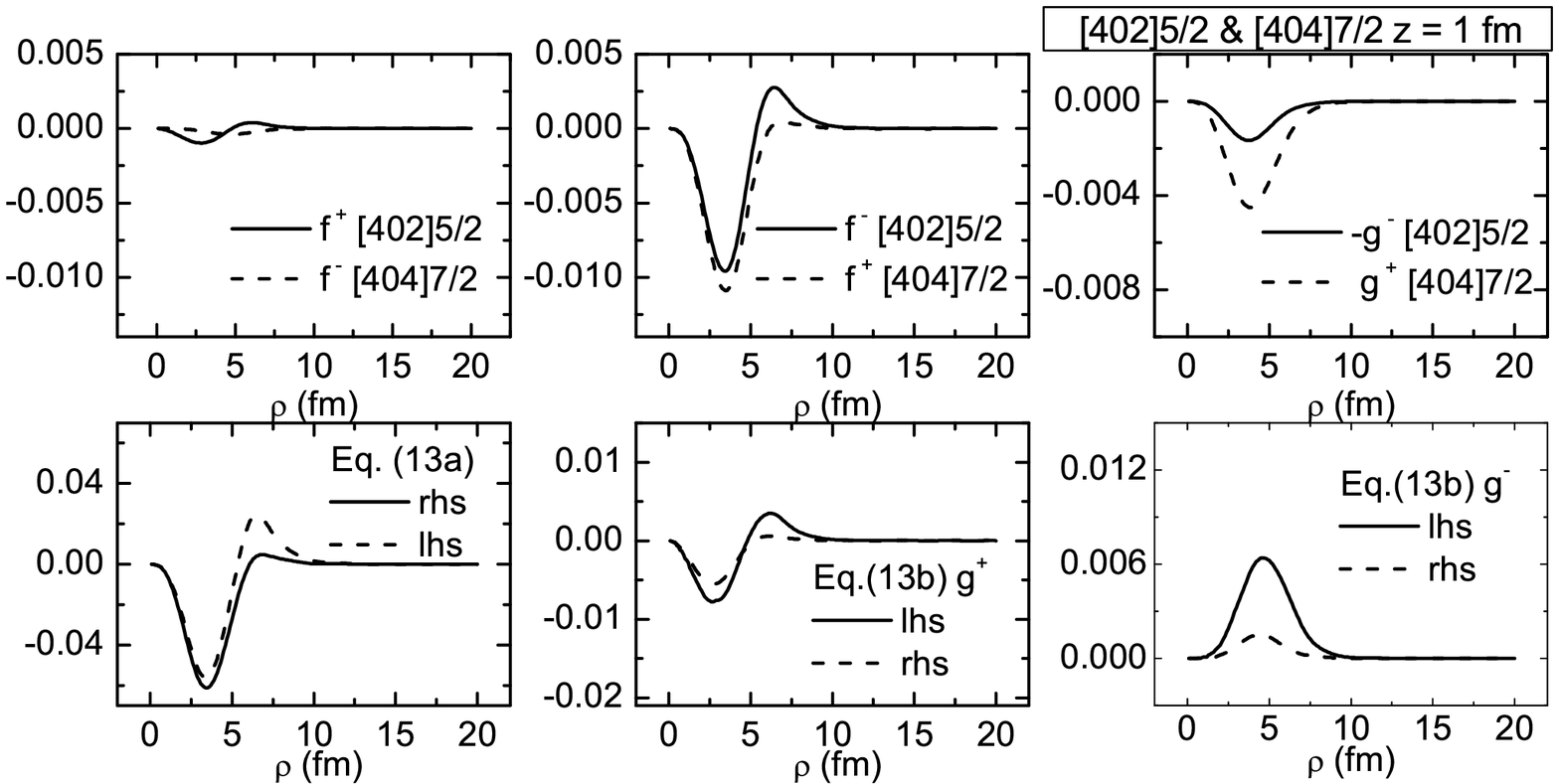}
\includegraphics[width=14.5cm]{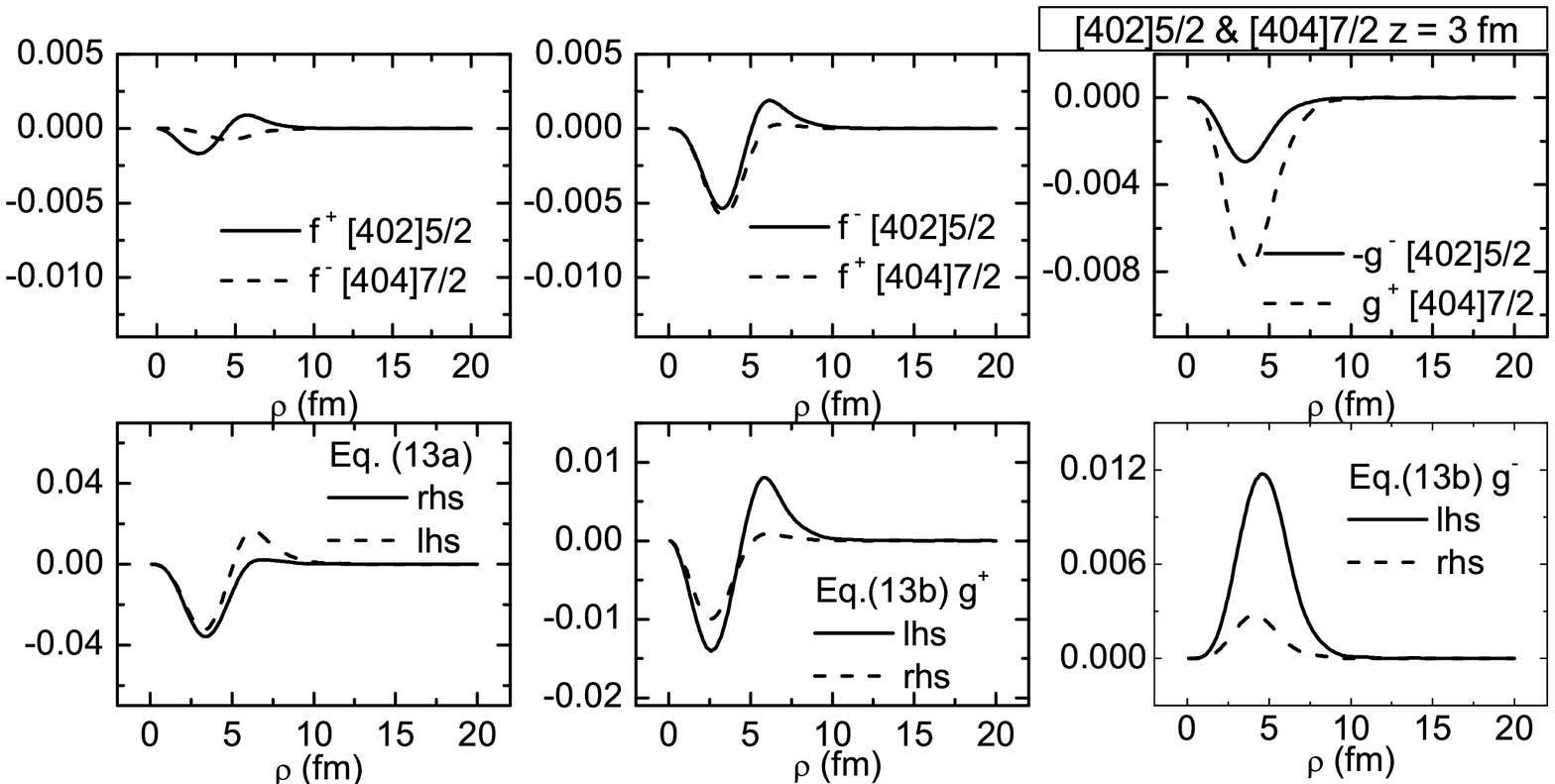}
\includegraphics[width=14.5cm]{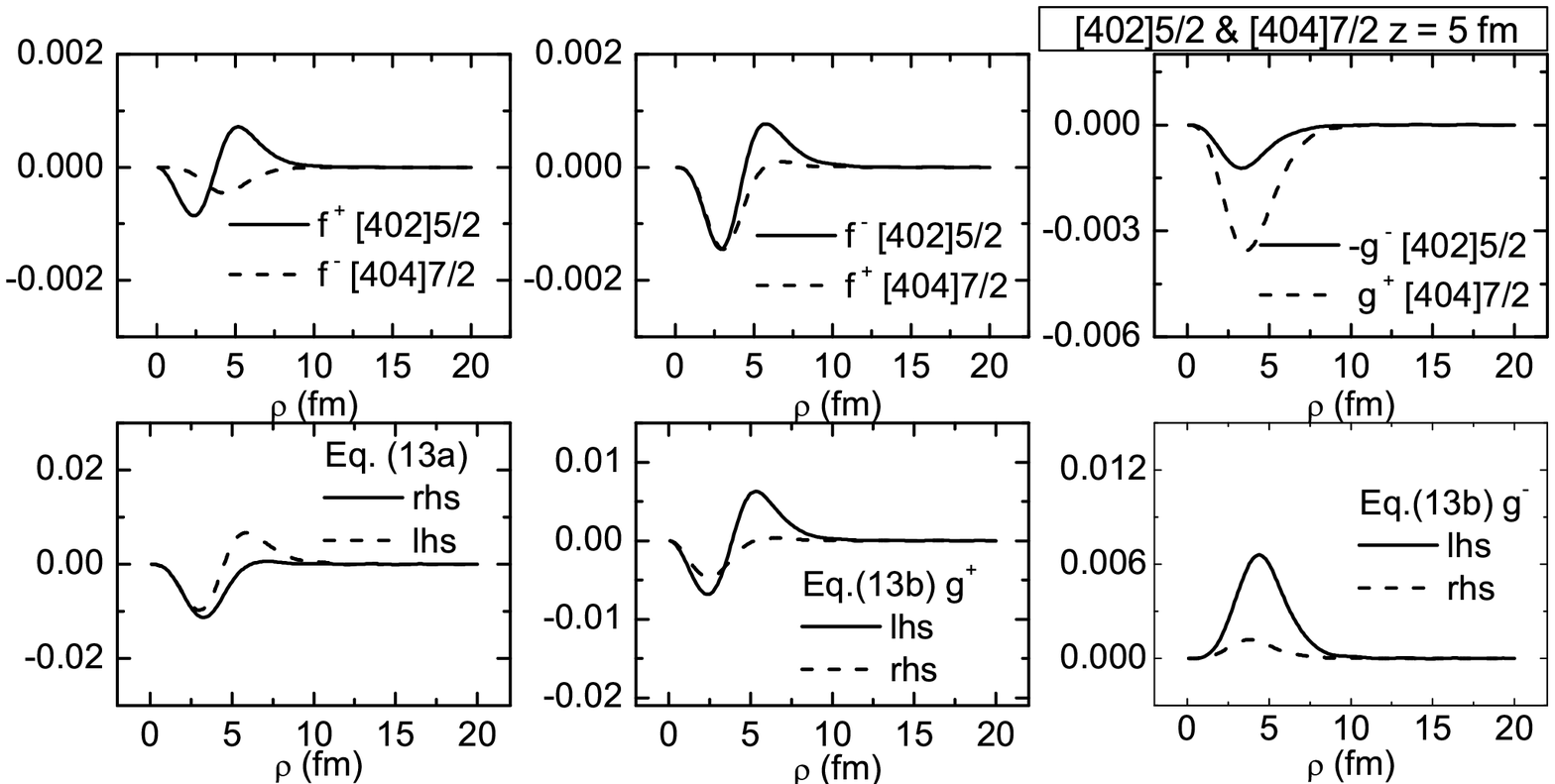}
\end{center}
\caption{Wave functions in (Fermi)$^{-3/2}$ as a function of $\rho$
and $z= 1,\, 3,\, 5$ fm for the neutron
pseudospin doublet [402]5/2 and [404]7/2 $({\tilde \Lambda} =3)$
in $^{168}$Er. The content of the graphs in each segment as
in Fig.~2.} \label{fig3}
\end{figure}

\begin{figure}
\begin{center}
\includegraphics[width=14.5cm]{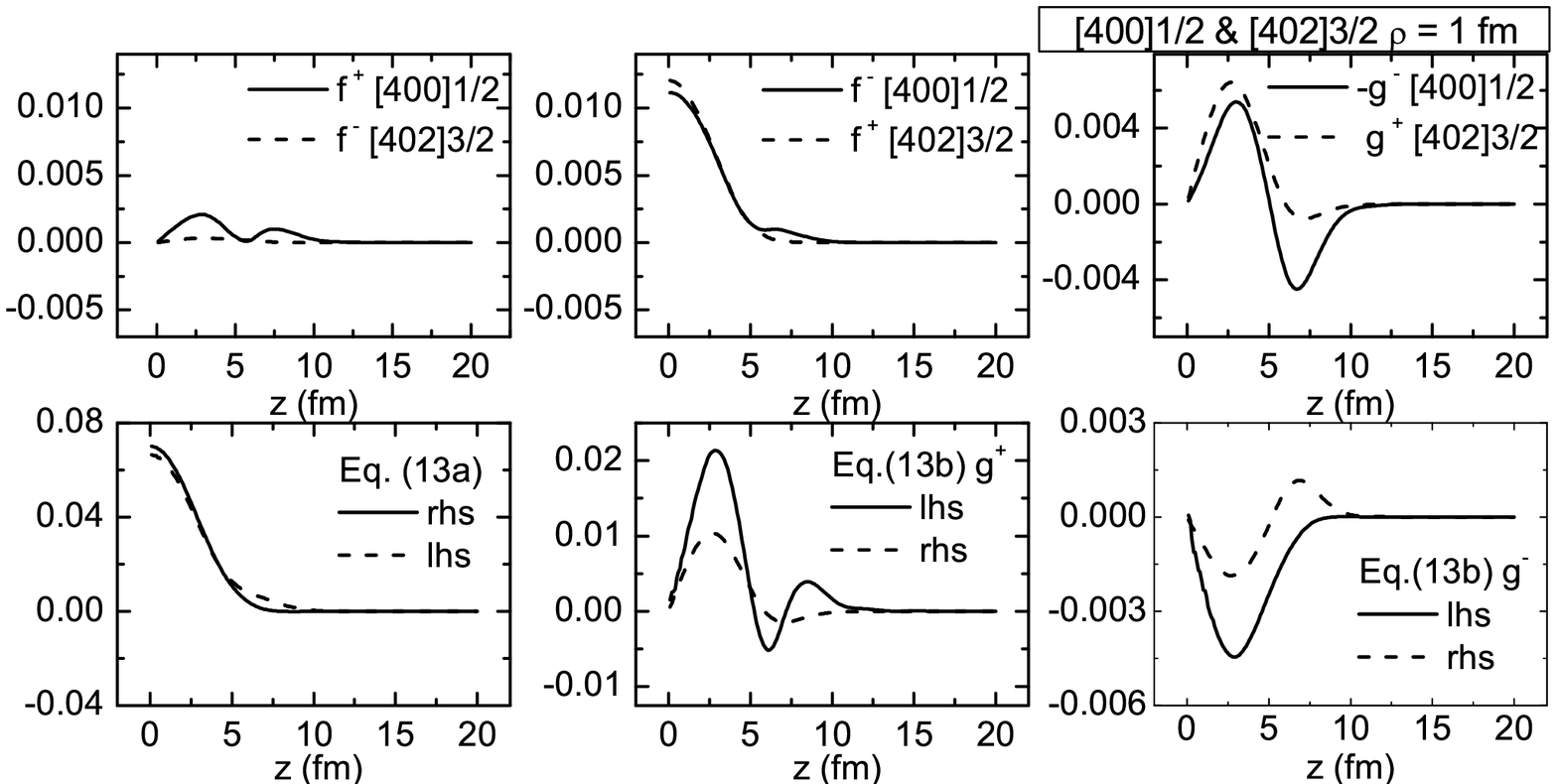}
\includegraphics[width=14.5cm]{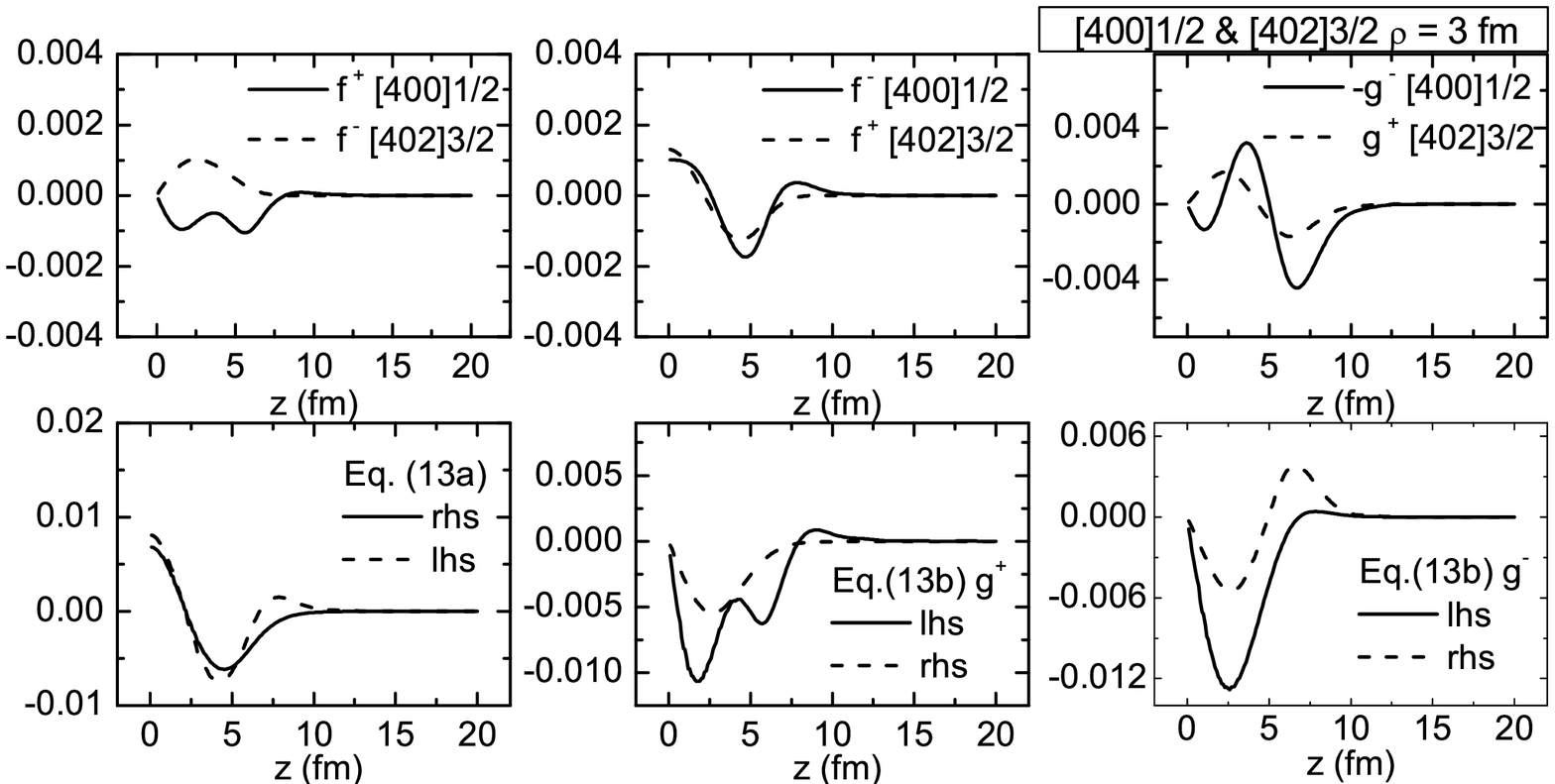}
\includegraphics[width=14.5cm]{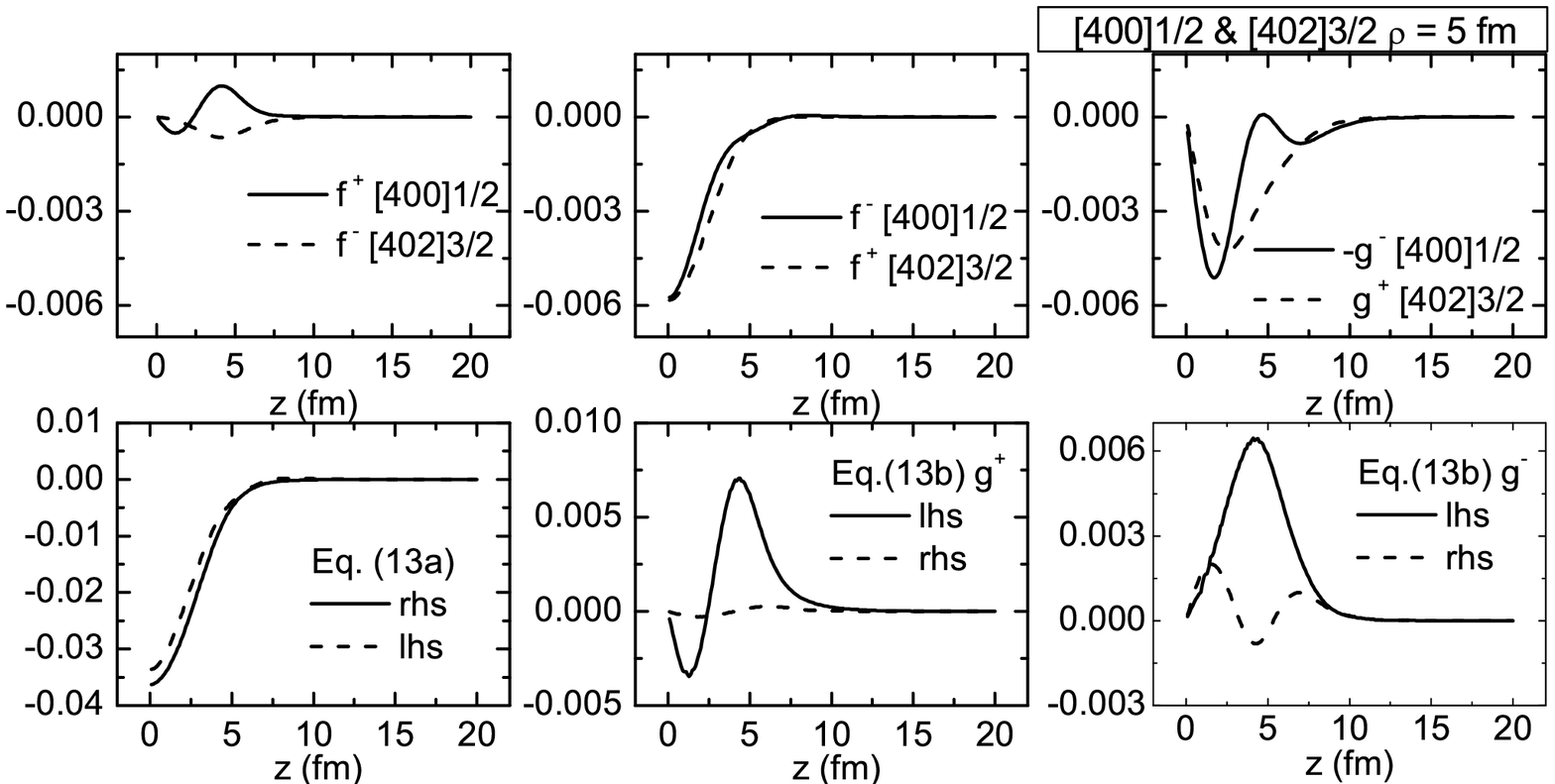}
\end{center}
\caption{ As in Fig. 2 but for the neutron pseudospin doublet
[400]1/2 and [402]3/2 (${\tilde \Lambda}$ =1) in
$^{168}$Er.}\label{fig4}
\end{figure}

\begin{figure}
\begin{center}
\includegraphics[width=14.5cm]{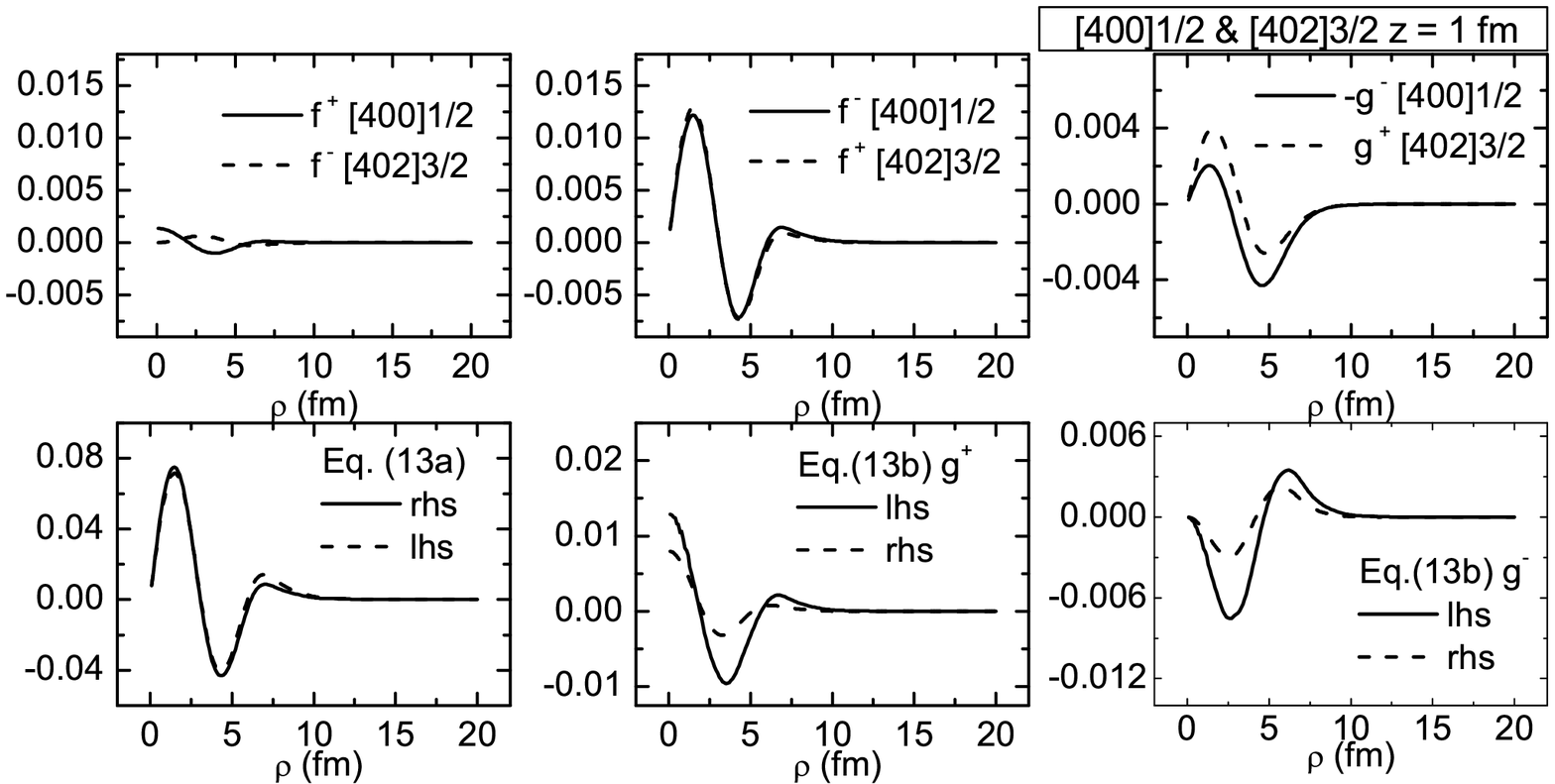}
\includegraphics[width=14.5cm]{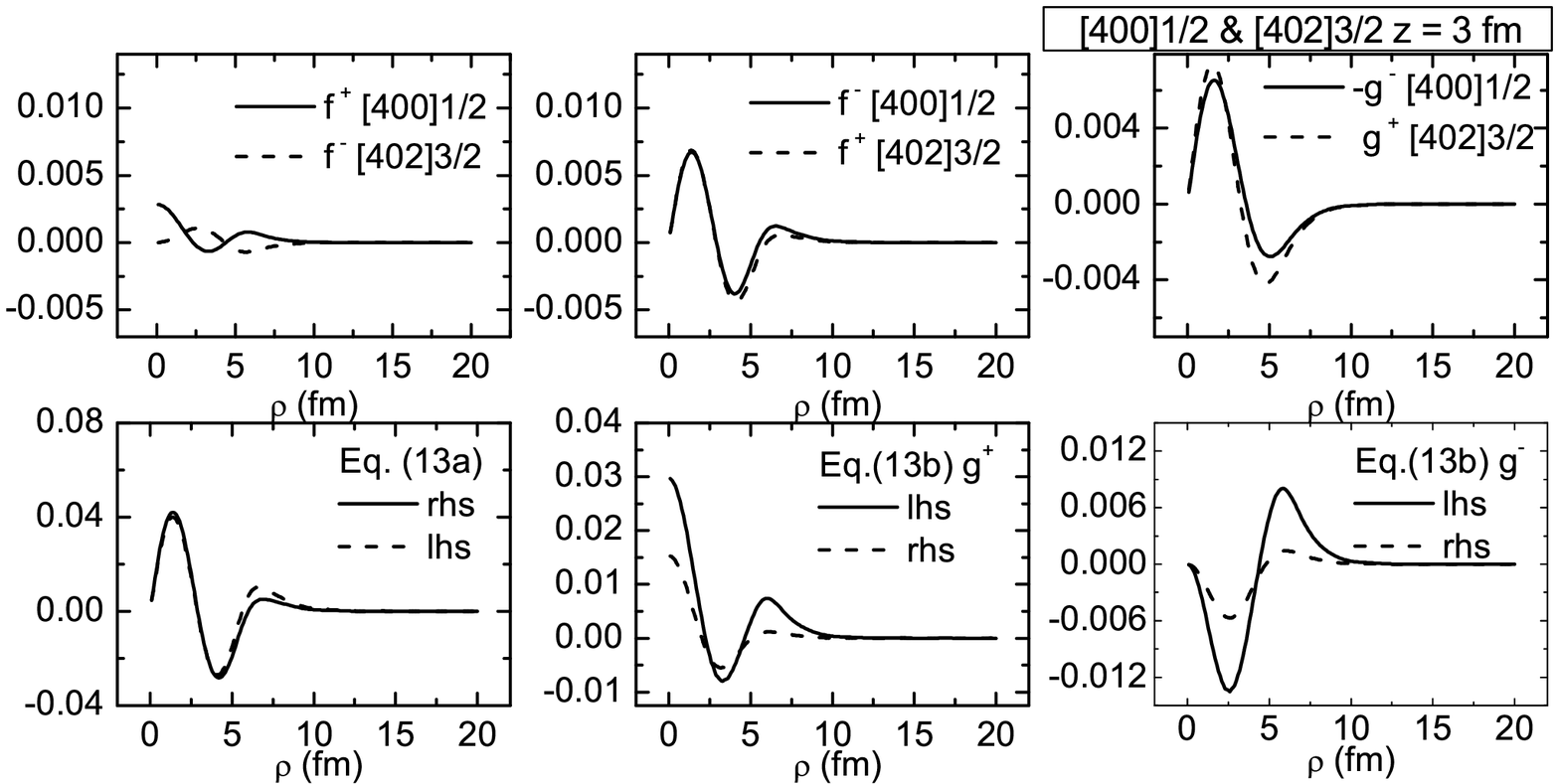}
\includegraphics[width=14.5cm]{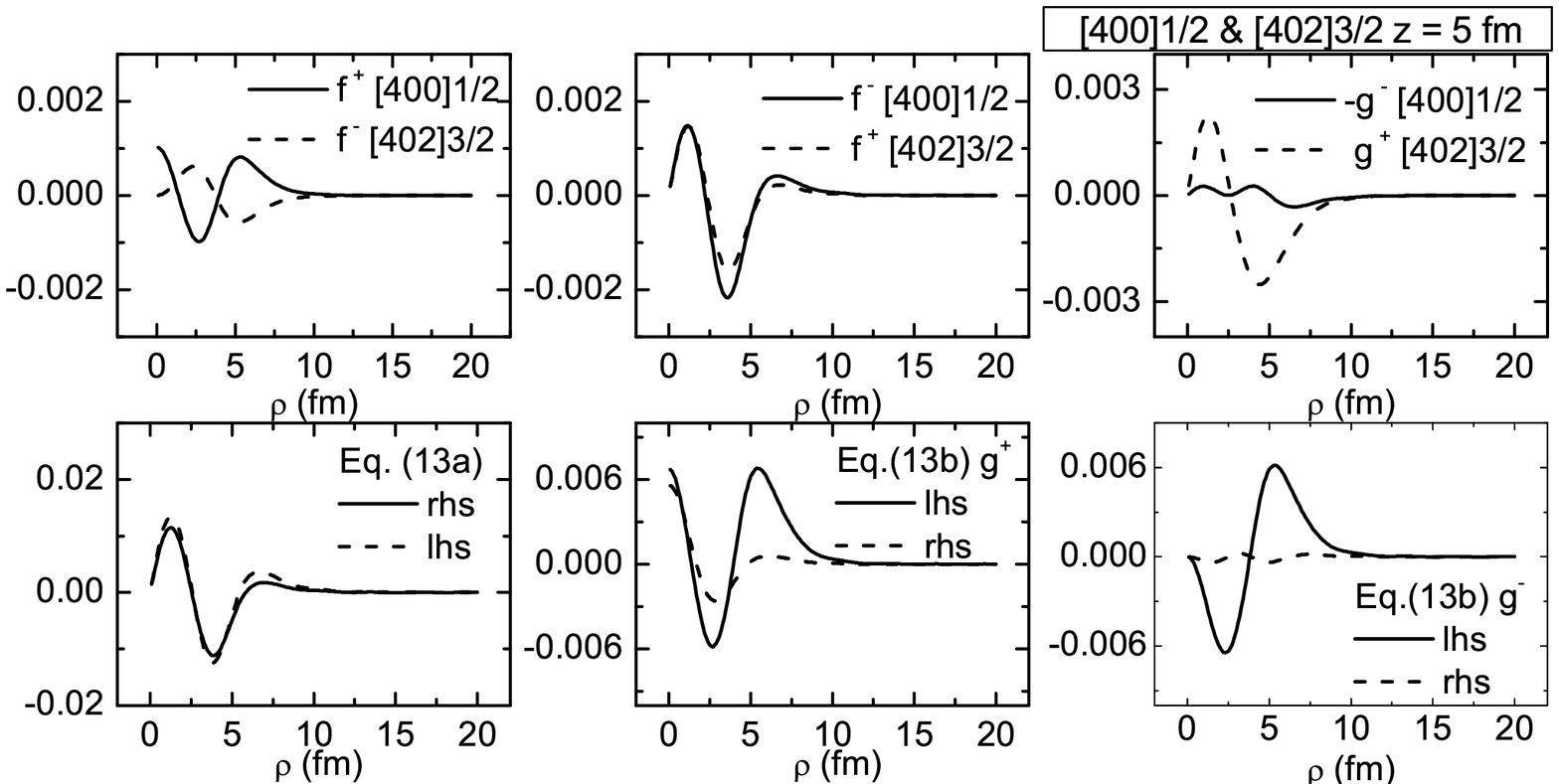}
\end{center}
\caption{ As in Fig. 3 but for the neutron pseudospin doublet
[400]1/2 and [402]3/2 (${\tilde \Lambda}$ =1) in
$^{168}$Er.}\label{fig5}
\end{figure}

\begin{figure}
\begin{center}
\includegraphics[width=14.5cm]{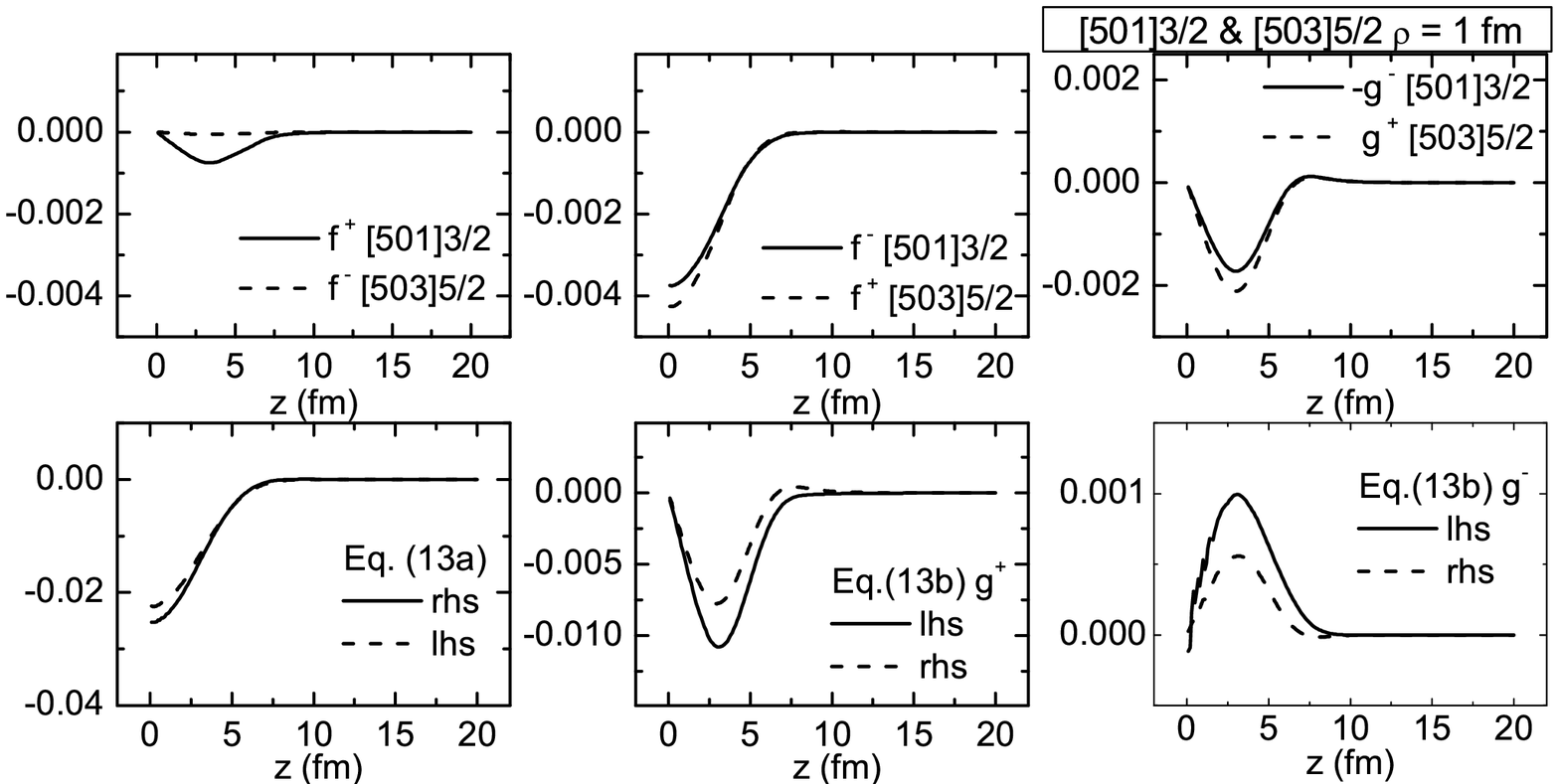}
\includegraphics[width=14.5cm]{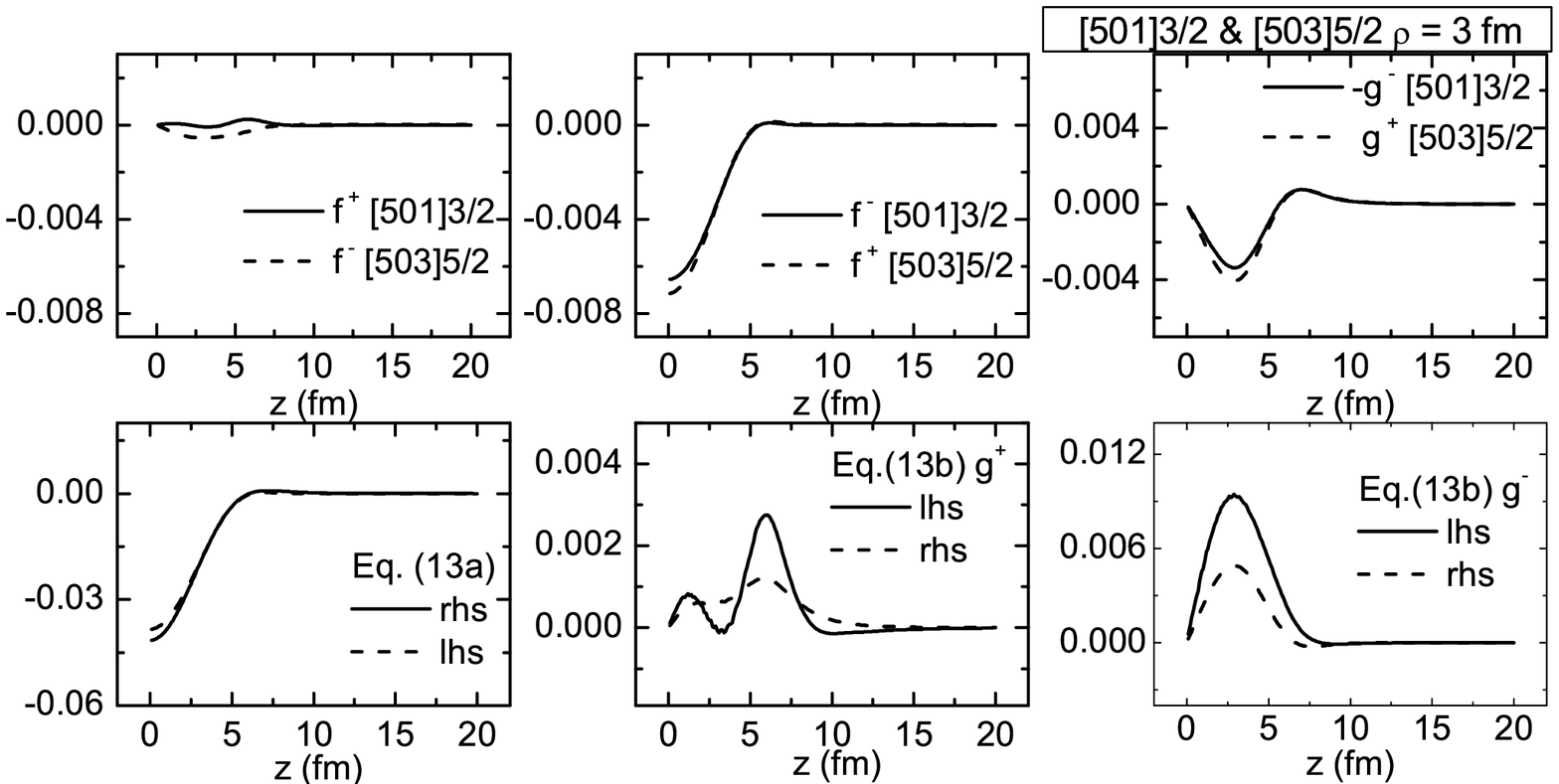}
\includegraphics[width=14.5cm]{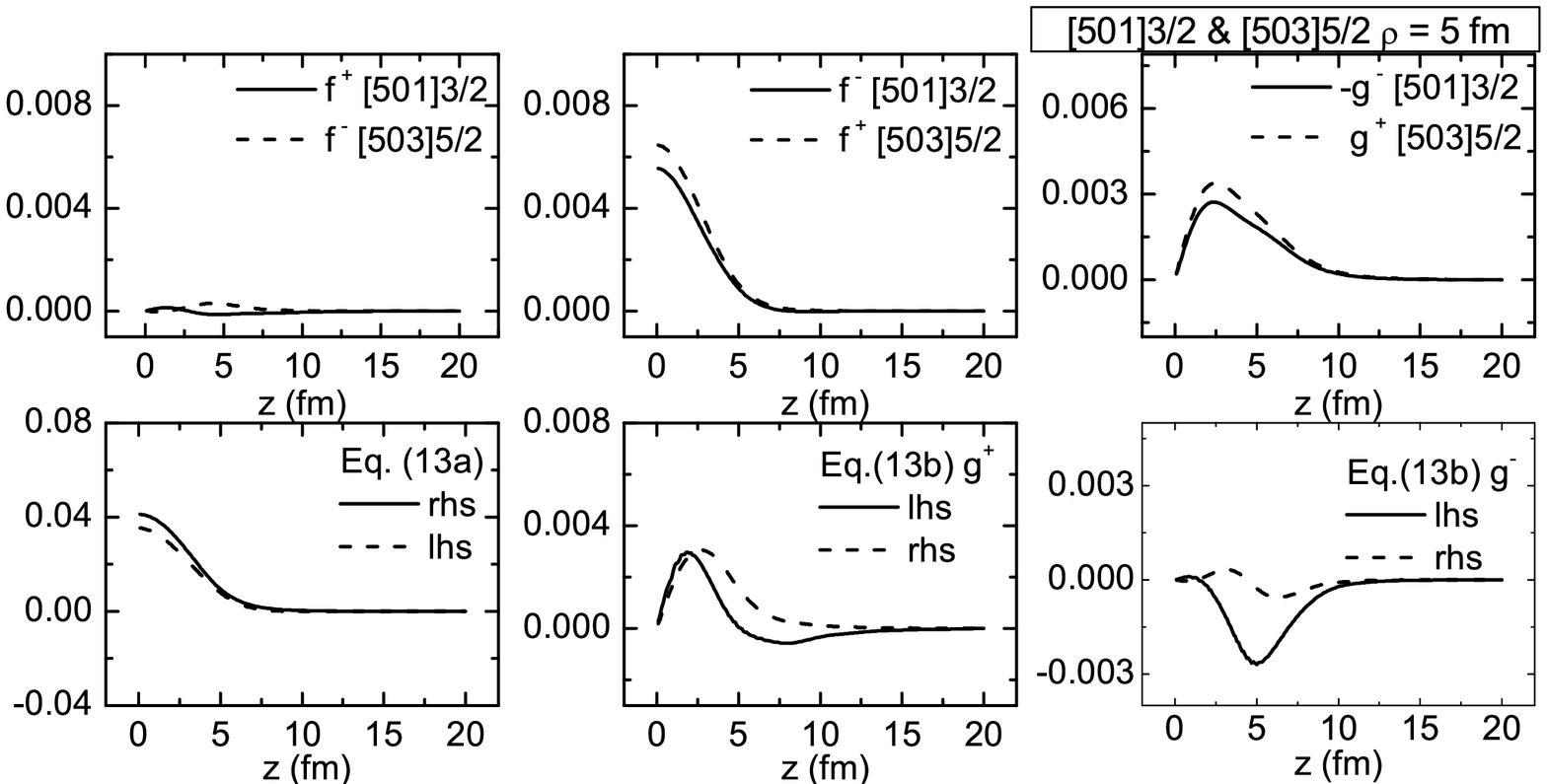}
\end{center}
\caption{ As in Fig. 2 but for the neutron pseudospin doublet
[501]3/2 and [503]5/2 (${\tilde \Lambda}$ =2) in
$^{168}$Er.}\label{fig6}
\end{figure}

\begin{figure}
\begin{center}
\includegraphics[width=14.5cm]{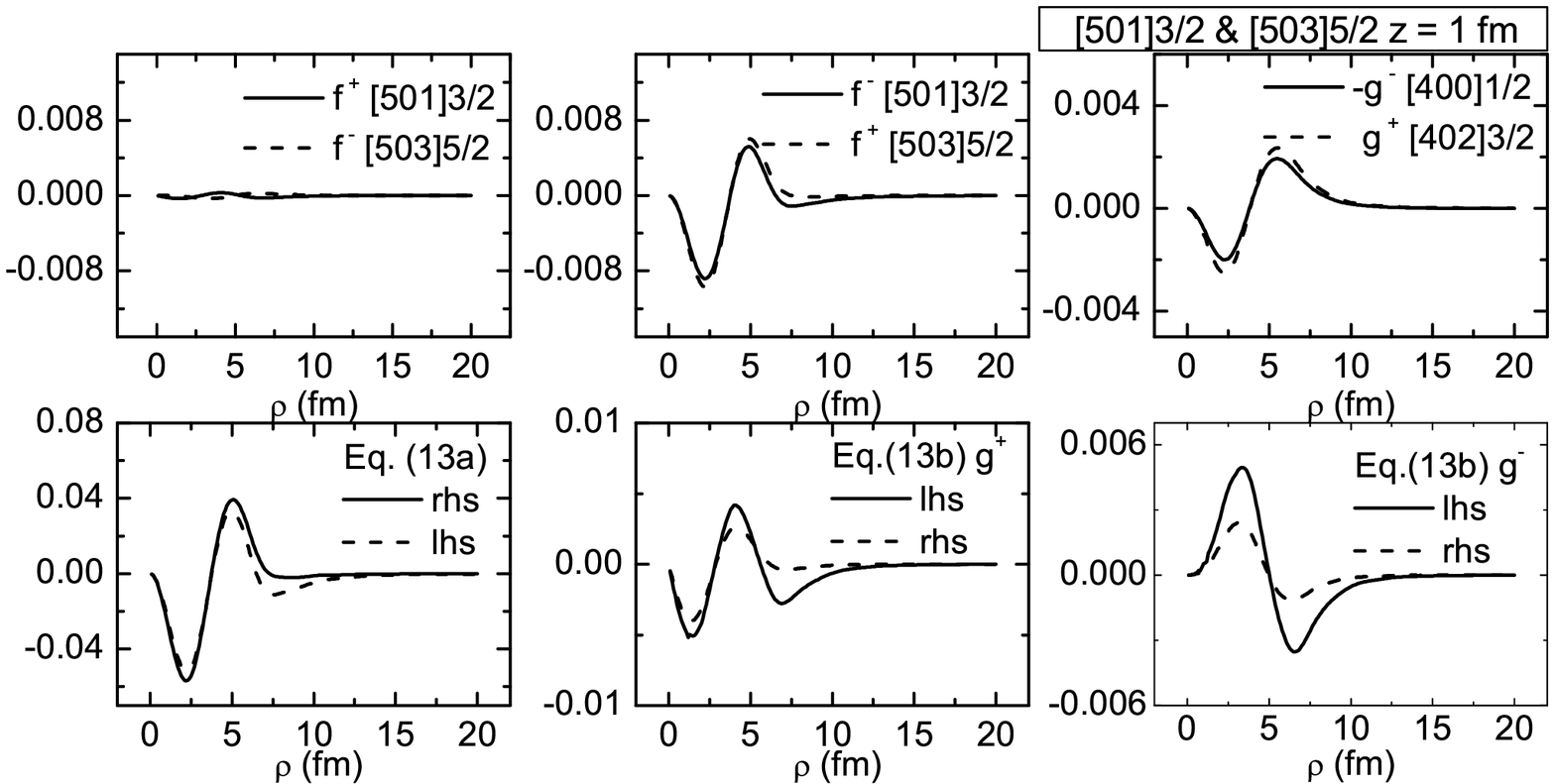}
\includegraphics[width=14.5cm]{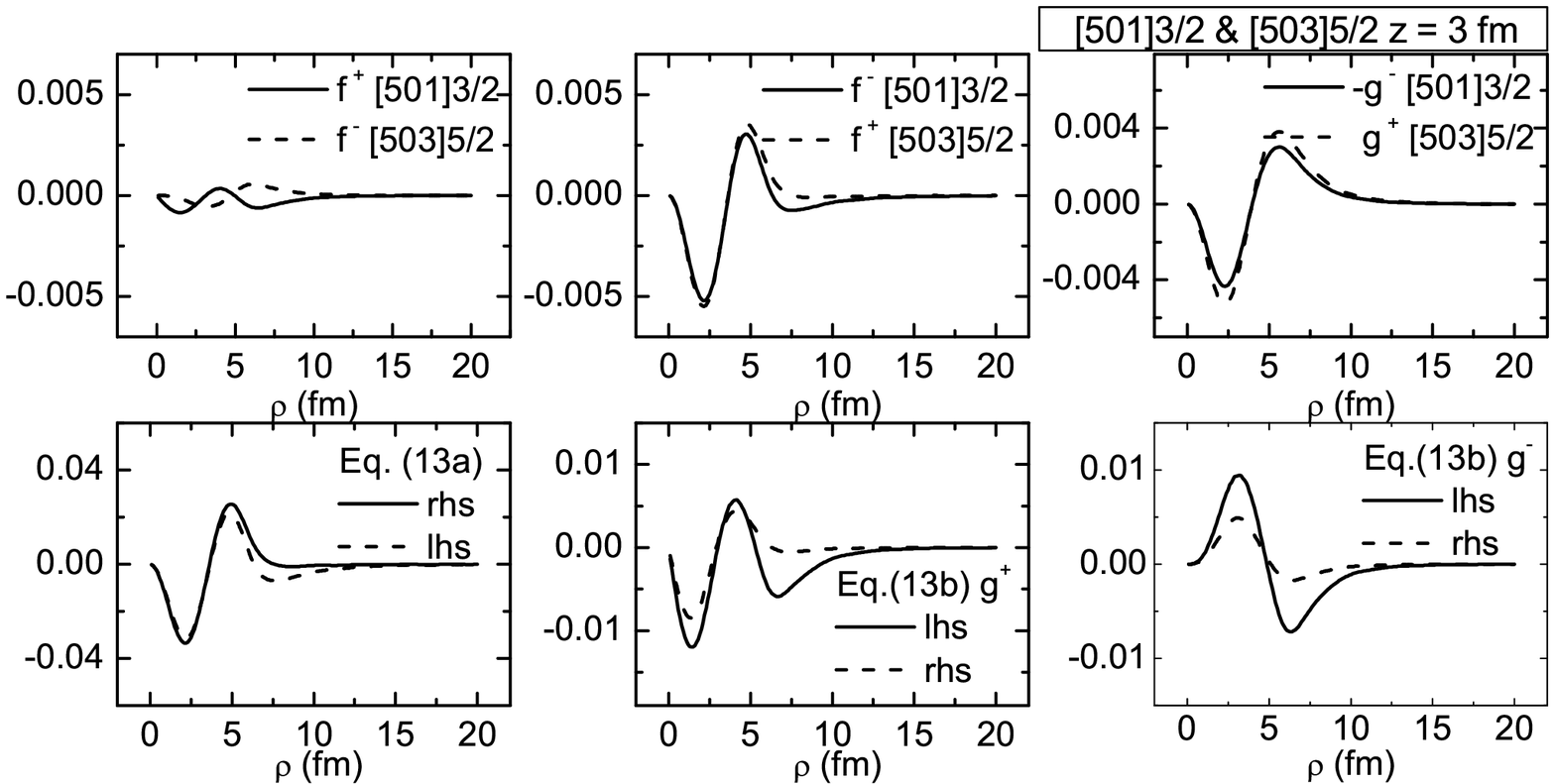}
\includegraphics[width=14.5cm]{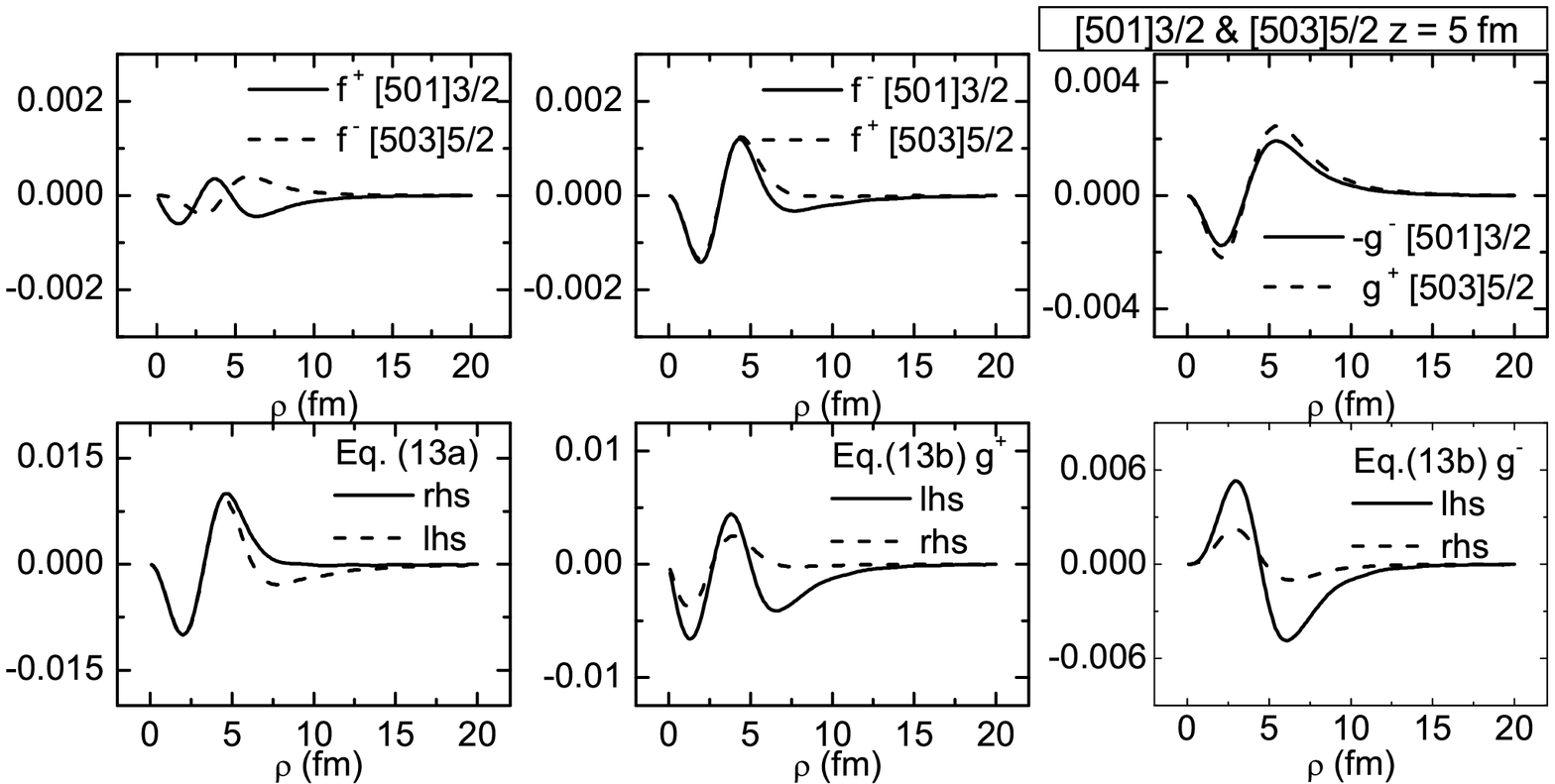}
\end{center}
\caption{ As in Fig. 3 but for the neutron pseudospin doublet
[501]3/2 and [503]5/2 (${\tilde \Lambda}$ =2) in
$^{168}$Er.}\label{fig7}
\end{figure}

\begin{figure}
\begin{center}
\includegraphics[width=14.5cm]{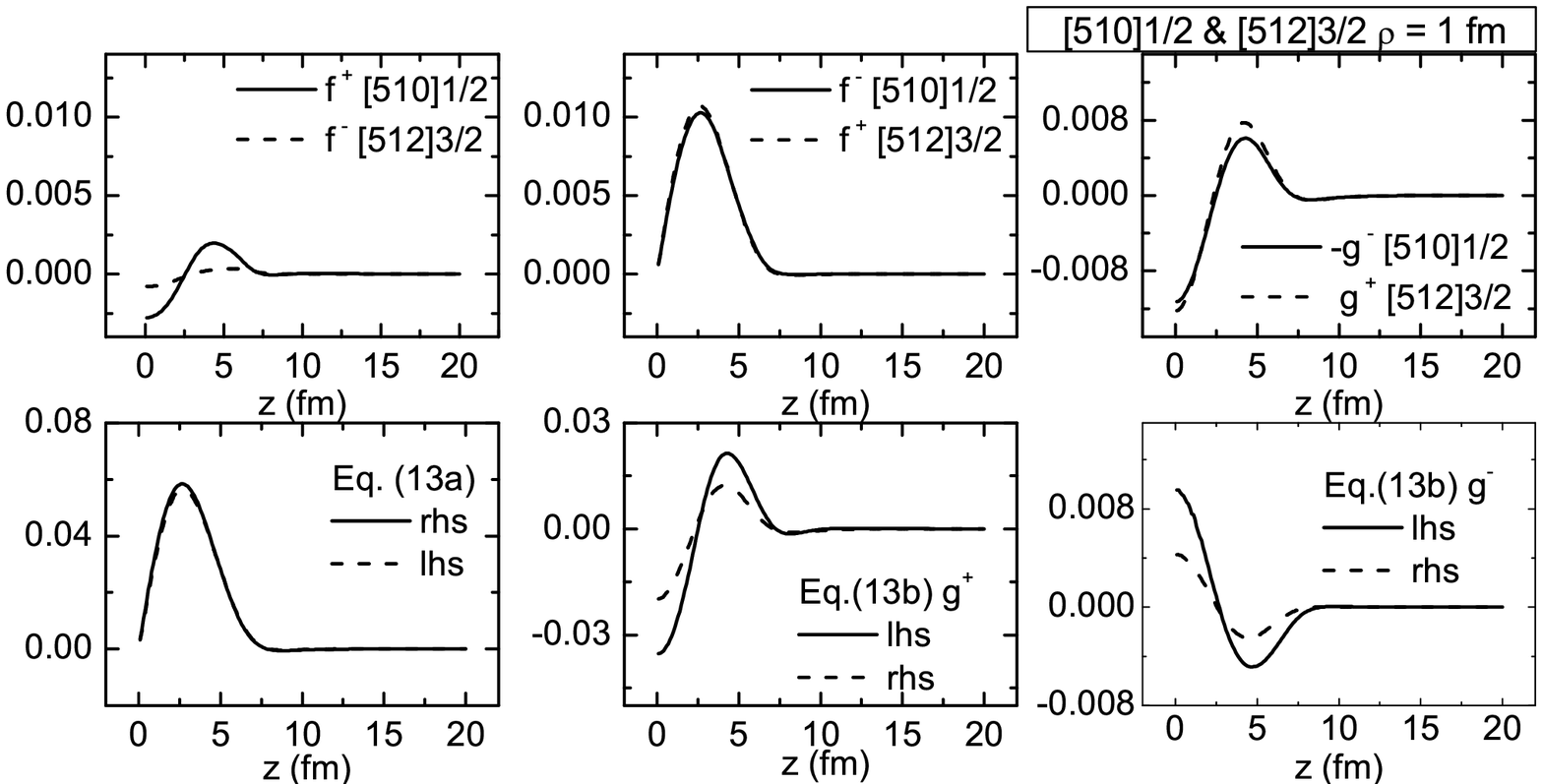}
\includegraphics[width=14.5cm]{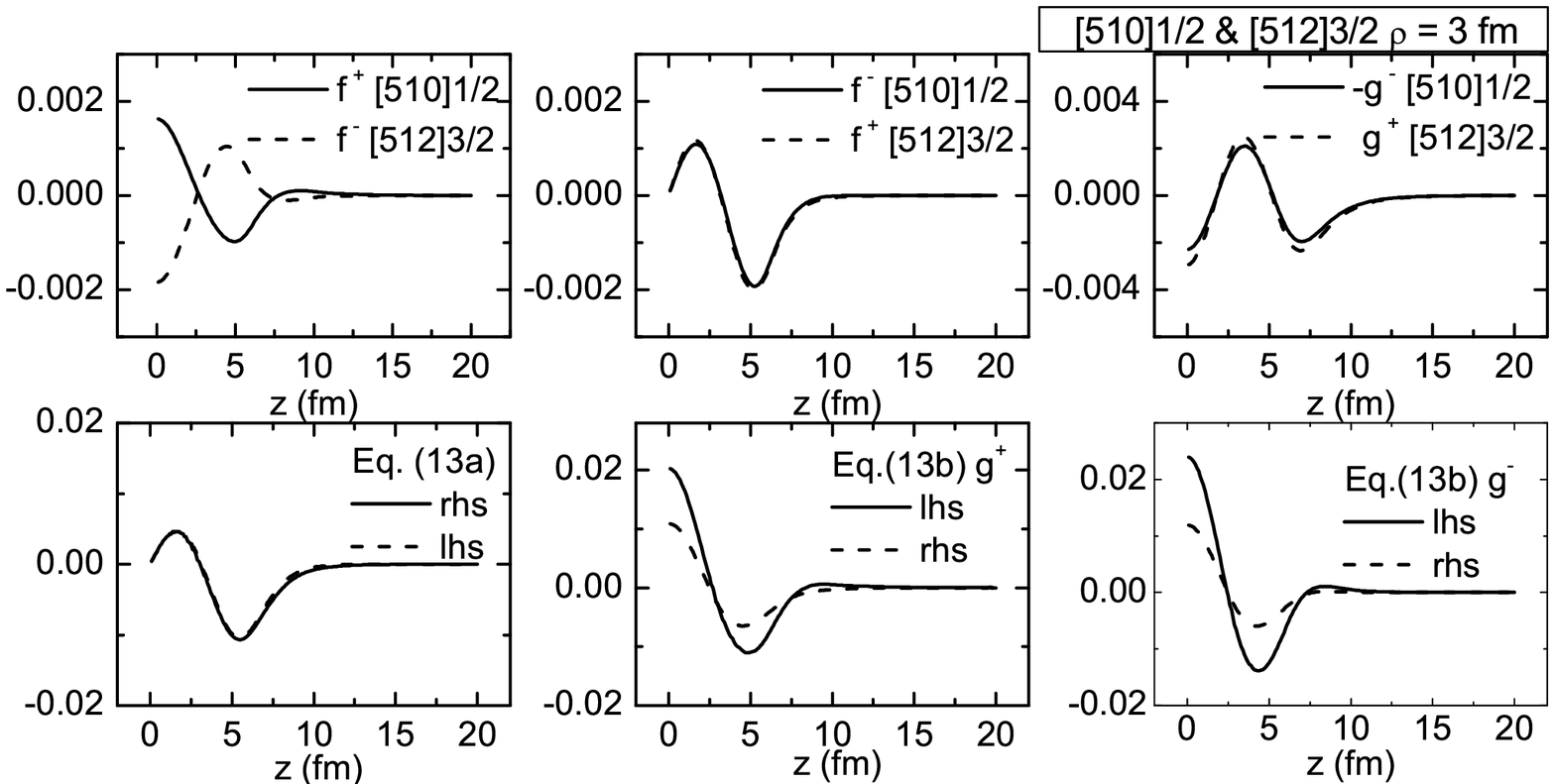}
\includegraphics[width=14.5cm]{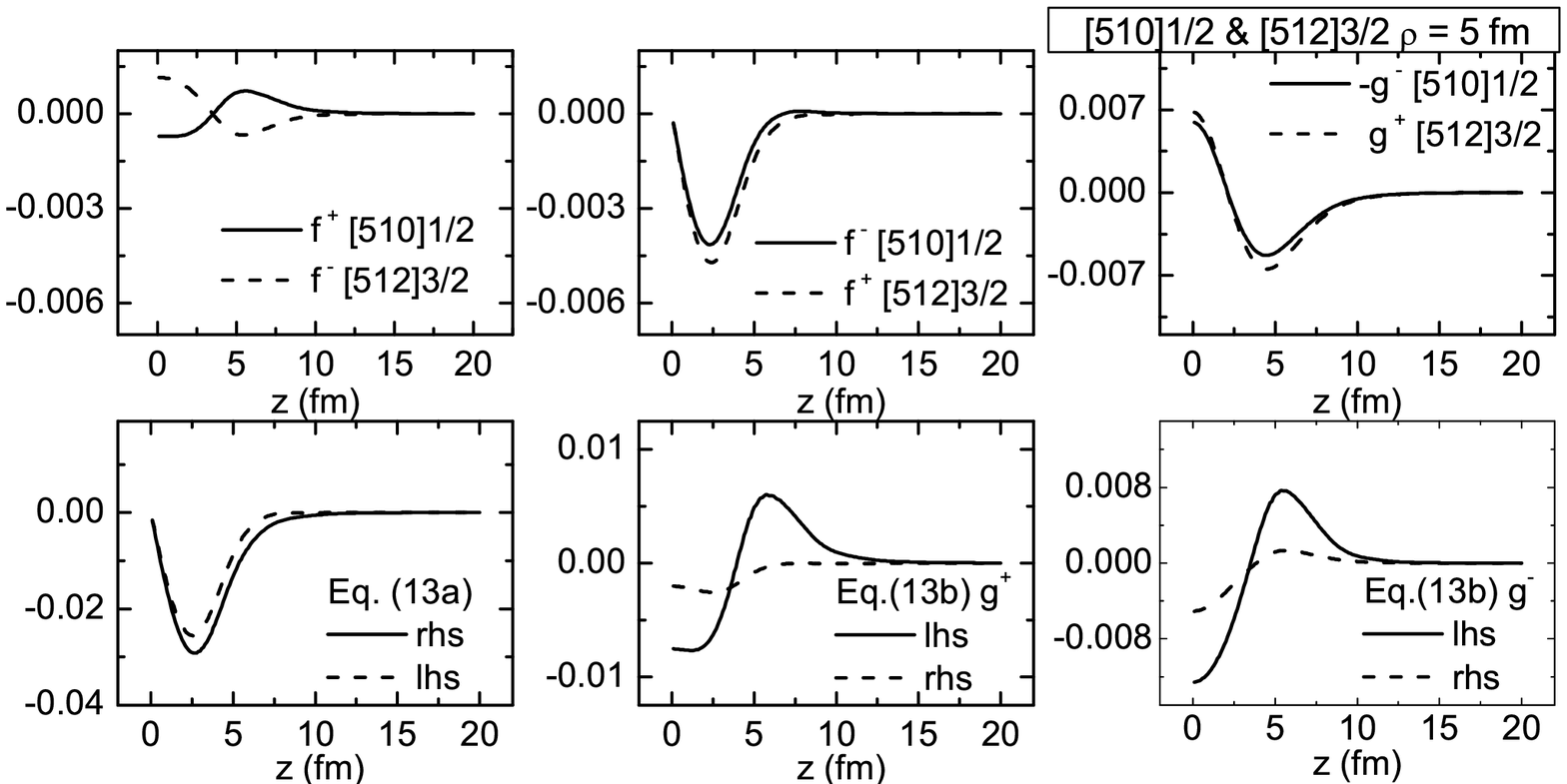}
\end{center}
\caption{ As in Fig. 2 but for the neutron pseudospin doublet
[510]1/2 and [512]3/2 (${\tilde \Lambda}$ =1) in
$^{168}$Er.}\label{fig8}
\end{figure}

\begin{figure}
\begin{center}
\includegraphics[width=14.5cm]{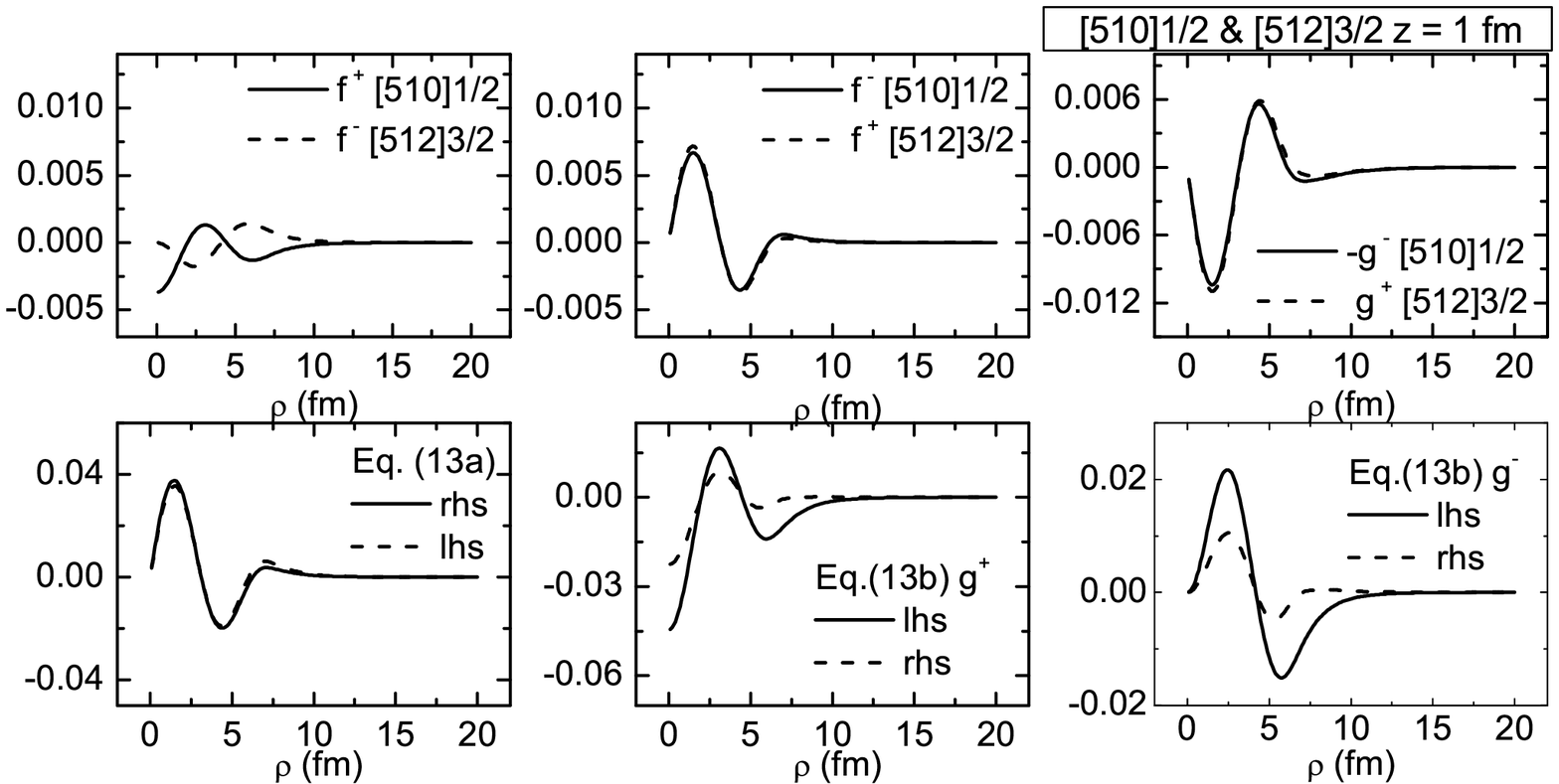}
\includegraphics[width=14.5cm]{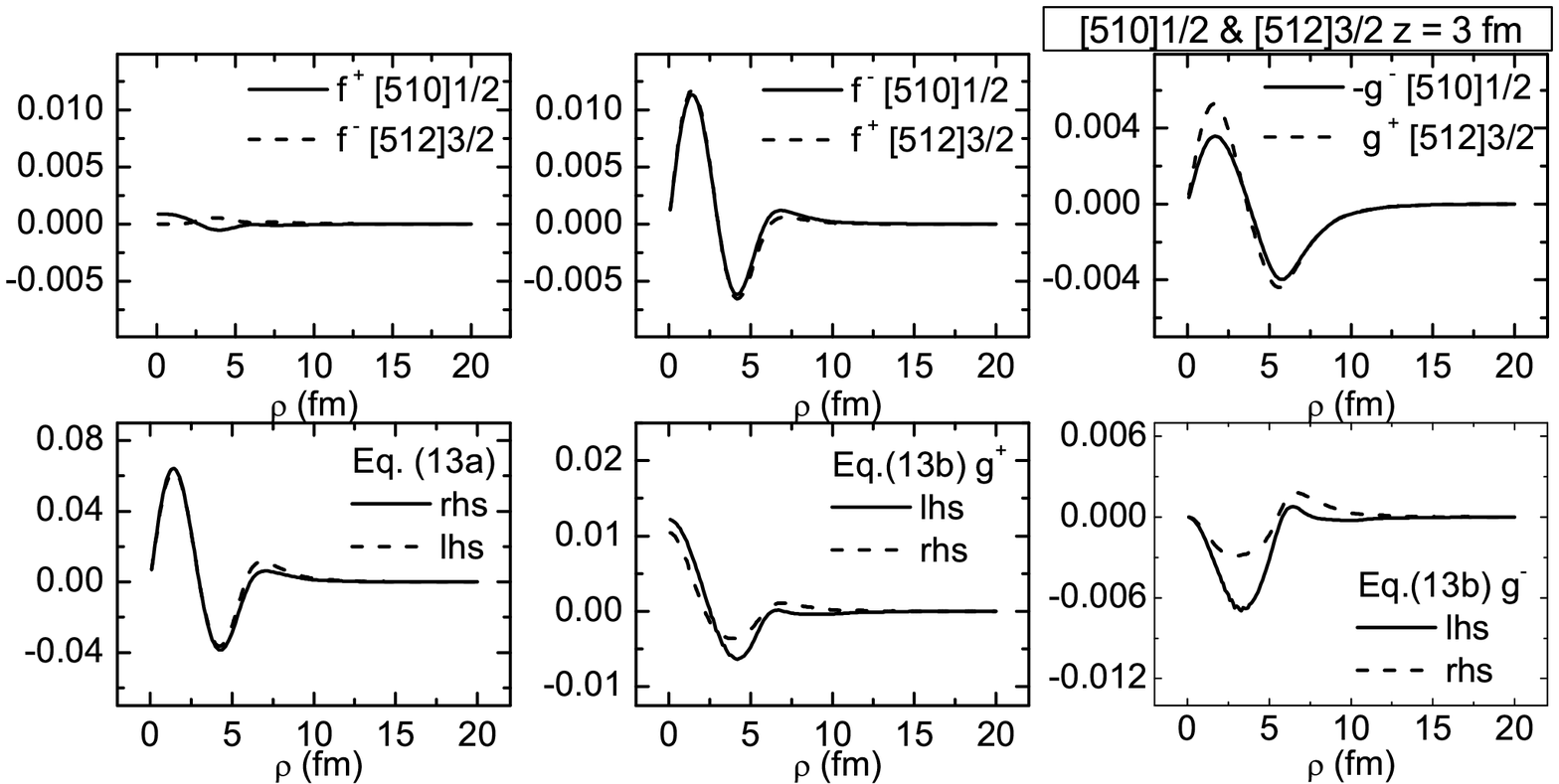}
\includegraphics[width=14.5cm]{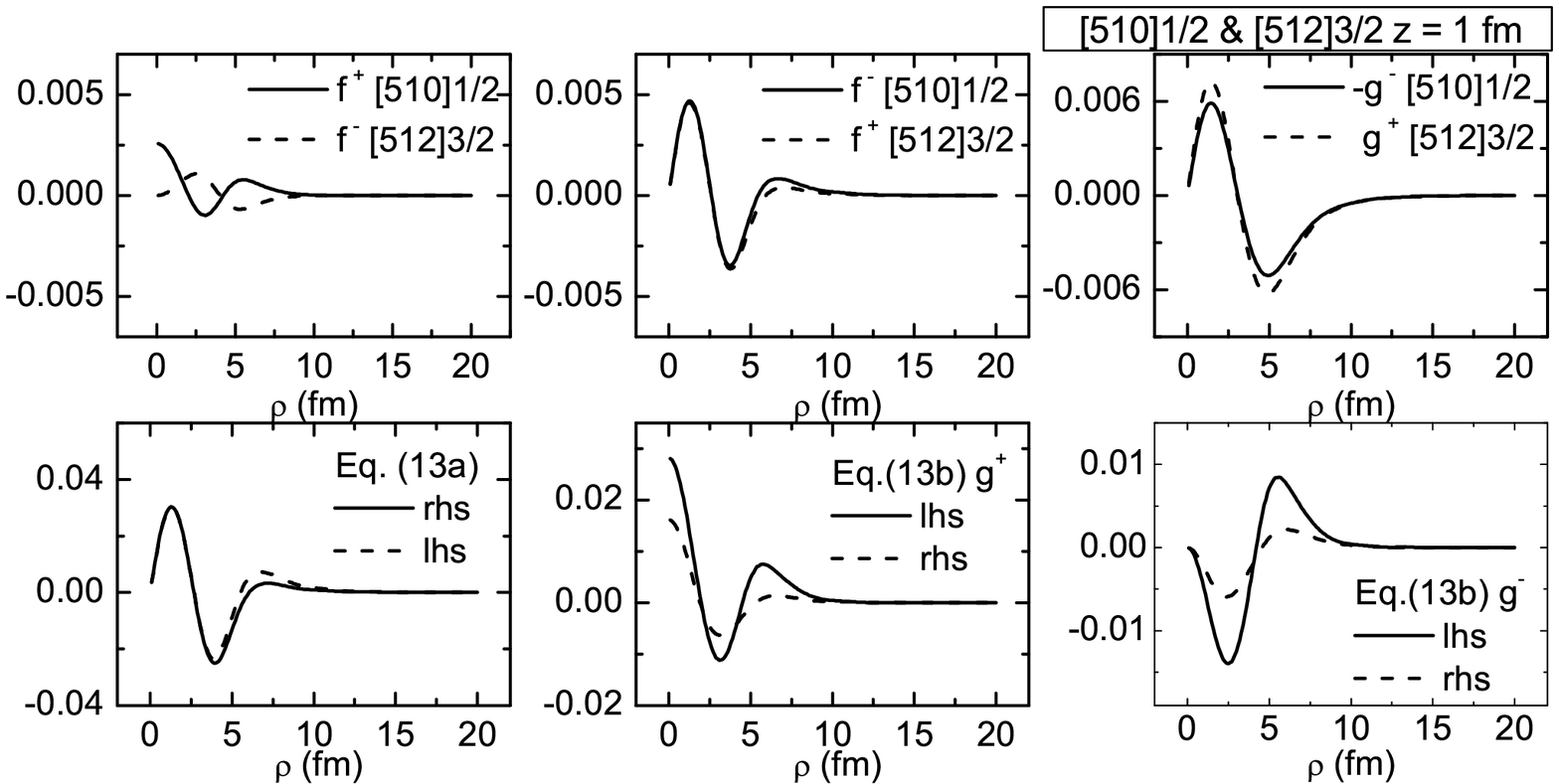}
\end{center}
\caption{ As in Fig. 3 but for the neutron pseudospin doublet
[510]1/2 and [512]3/2 (${\tilde \Lambda}$ =1) in
$^{168}$Er.}\label{fig9}
\end{figure}

\end{document}